\documentclass[11pt]{article}

\usepackage[latin1]{inputenc}
\usepackage{amstext,amsmath,amssymb,amsfonts,bbm,amsmath}
\usepackage{amssymb}
\usepackage{extarrows}
\usepackage{xcolor} 
\usepackage[colorlinks, linkcolor=blue, citecolor=blue, urlcolor=blue]{hyperref}
\usepackage{bm}
\usepackage{authblk}
\usepackage{caption} 
\usepackage{epsfig} 
\usepackage{xcolor} 
\usepackage{graphicx} 
\usepackage{braket} 
\usepackage{slashed} 
\usepackage{dsfont} 
\usepackage{amsthm}  

\usepackage{cite}

\allowdisplaybreaks[4]

\usepackage{psfrag}
\usepackage{tikz} 
\usetikzlibrary{arrows}
\usetikzlibrary{plotmarks}
\usetikzlibrary{external}
\usetikzlibrary{patterns}
\usepackage{pgfplots}
\pgfplotsset{compat=1.9}
\usetikzlibrary{patterns}
\usetikzlibrary{calc}
\usetikzlibrary{automata,positioning}

\usepackage{float}  
\usepackage{subfig}
\usepackage[lmargin=50pt,rmargin=60pt,tmargin=60pt,bmargin=65pt]{geometry}

\DeclareMathOperator{\Tr}{Tr}

\newcommand{\be}{\begin{equation}}
	\newcommand{\ee}{\end{equation}} 

\newcommand{\f}{\frac}
\newcommand{\p}{\partial}
\newcommand{\la}{\langle}
\newcommand{\ra}{\rangle}
\newcommand{\dd}{{\rm d}^d}
\newcommand{\pright}{\overset{\scriptscriptstyle\rightarrow}{\partial}}
\newcommand{\pleft}{\overset{\scriptscriptstyle\leftarrow}{\partial}}

\let\a=\alpha \let\b=\beta  \let\g=\gamma  \let\d=\delta 
\let\z=\zeta    \let\l=\lambda
\let\m=\mu    \let\n=\nu      \let\r=\rho 
\let\s=\sigma \let\t=\tau   
\let\G=\Gamma \let\D=\Delta   \let\L=\Lambda 
\let\Om=\Omega  \let\eps=\epsilon

\newcommand{\cA}{\mathcal{A}}

\newcommand{\cC}{\mathcal{C}}

\newcommand{\cL}{\mathcal{L}}

\newcommand{\cO}{\mathcal{O}}

\newcommand{\cT}{\mathcal{T}}

\newcommand{\mbp}{\mathbf{p}}
\newcommand{\mbx}{\mathbf{x}}

\allowdisplaybreaks[4]

\numberwithin{equation}{section}

\theoremstyle{remark}

\begin{document}

	\title{\bf Finite-size versus finite-temperature effects\\ in the critical long-range $O(N)$ model}
	
	\author[1]{Dario Benedetti}
 	\author[2]{Razvan Gurau}
  \author[3]{Sabine Harribey}
  \author[2]{Davide Lettera}
	
\affil[1]{\normalsize\it 
CPHT, CNRS, \'Ecole polytechnique, Institut Polytechnique de Paris, 91120 Palaiseau, France
\authorcr \hfill }
	
	\affil[2]{\normalsize\it 
		Heidelberg University, Institut f\"ur Theoretische Physik, Philosophenweg 19, 69120 Heidelberg, Germany
		\authorcr \hfill
		}

\affil[3]{\normalsize \it 
 NORDITA, Stockholm University and KTH Royal Institute of Technology,\authorcr Hannes Alfv\'{e}ns v\"{a}g 12, SE-106 91 Stockholm, Sweden
  \authorcr \hfill
  \authorcr
  emails: dario.benedetti@polytechnique.edu, gurau@thphys.uni-heidelberg.de, sabine.harribey@su.se, lettera@thphys.uni-heidelberg.de 
 \authorcr \hfill }
 
	\date{}
	
	\maketitle
	
	\hrule\bigskip
	
	\begin{abstract}

In this paper we consider classical and quantum versions of the critical long-range $O(N)$ model, for which we study finite-size and finite-temperature effects, respectively, at large $N$. First, we consider the classical (isotropic) model, which is conformally invariant at criticality, and we introduce one compact spatial direction. We show that the finite size dynamically induces an effective mass and we compute the one-point functions for bilinear primary operators with arbitrary spin and twist. Second, we study the quantum model, mapped to a Euclidean anisotropic field theory, local in Euclidean time and long-range in space, which we dub \emph{fractional Lifshitz field theory}. 
We show that this model admits a fixed point at zero temperature, where it displays anisotropic Lifshitz scaling, and show that at finite temperature a thermal mass is induced. We then compute the one-point functions for an infinite family of bilinear scaling operators. 

In both the classical and quantum model, we find that, as previously noted for the short-range $O(N)$ model in \cite{Iliesiu:2018fao}, the large-$N$ two-point function contains information about the one-point functions, not only of the bilinear operators, but also of operators that appear in the operator product expansion of two fundamental fields only at subleading order in $1/N$, namely powers of the Hubbard-Stratonovich intermediate field.

	\end{abstract}
	
	\hrule\bigskip
	
	\tableofcontents
	
\section{Introduction}

Quantum field theories in high energy physics are typically based on Lagrangians which only depend on the fields and their first derivatives.
Higher-order derivatives in time are excluded in order to preserve unitarity, and then by Lorentz invariance higher-order derivatives in space are excluded as well. 
Fractional derivatives of order less than one are instead excluded by locality, or the need of having a particle interpretation.
Such requirements might be dropped in Euclidean field theories, if the base space is interpreted as representing only spatial directions and the theory itself as a model of classical equilibrium statistical physics.
It is within this perspective that we can frame the study of non-local models containing fractional derivatives, also known as long-range models.
Such models have been realized experimentally in recent years (see \cite{Defenu:2021glw} for a review), but have been studied from a theoretical standpoint since a long time because of several interesting features: they can violate the Mermin-Wagner theorem and admit phase transitions even in one dimension, as proved by Dyson for the long-range Ising model \cite{Dyson:1968up}; they typically lead to one-parameter families of universality classes (as found for example by two-loop  \cite{Fisher:1972zz,Yamazaki:1977pt} and three-loop computations \cite{Benedetti:2020rrq,Behan:2023ile}) and tuning such parameter one can study, at fixed dimension, phenomena such as the crossover from a long-range to a short-range universality class \cite{Sak:1973,Brezin:2014,Behan:2017dwr,Behan:2017emf}; one can rigorously construct fixed points and trajectories of the renormalization group in three dimensions \cite{Brydges:2002wq,Abdesselam:2006qg,Slade:2017,Lohmann:2017,Giuliani:2020aot}; and so on.
Their realizations in the context of defect or boundary field theories have also contributed to renewing theoretical interest in such models \cite{Paulos:2015jfa,Giombi:2019enr}.

In this paper, we want to address some aspects of finite-size and finite-temperature effects, which in general are very important (see for example \cite{Cardy:1988ag}), in  long-range models.
In particular, we will study the long-range $O(N)$ model with one compact direction.
This can be seen as a special case of finite-size effects in the corresponding classical statistical model \cite{Linhares:2012}.
However, by far the most common motivation for introducing only one compact direction is the fact that the quantum statistical mechanics of a $d$-dimensional system at finite temperature $T=1/\b$ can be described as a $(d+1)$-dimensional Euclidean field theory with compact Euclidean time of length $\b$, that is, on a spacetime $S^1_{\beta}\times\mathbb{R}^{d}$ with Euclidean signature (e.g.\ \cite{Bellac:2011kqa}).
Such a description has two complementary limits: the $\b\to \infty$ limit describing a quantum system at zero-temperature as a Euclidean field theory in $d+1$ dimensions, and the  $\b\to 0$ limit describing a classical system at equilibrium in $d$ space dimensions.
Therefore, in general with a given Euclidean field theory comes a choice of interpreting it as representing either a                classical or a quantum statistical system. 
However, this is not always the case, as the quantum interpretation is more constrained: for example,  unitarity implies that derivatives of order higher than two cannot be assigned to an imaginary-time direction.
Moreover, fractional derivatives should also be interpreted with care, as the reconstruction of a quantum Hamiltonian from a Lagrangian containing fractional time derivatives poses some challenges, and it might describe a system very different from  that associated to the classical interpretation. 
We stress this point because of our focus on long-range models, where the fact that finite-temperature and finite-size cases are structurally different seems to be sometimes overlooked. As we will discuss below, the finite-size classical system involves a compact direction among those appearing in the non-local kernel, while the field theoretic description of the finite-temperature quantum system requires the introduction of one extra compact direction with local interaction.
This observation is not really new, but with few exceptions \cite{Dutta:2001,Defenu:2017} it is often only left implicit in the details of numerical simulations \cite{Lazo:2021,zhao2023finitetemperature}, and it has not been explored much at an analytical level.

In order to clarify this point, let us consider a finite but continuous system\footnote{See \cite{Suzuki:1976cqdual} for a similar story in the case of discrete spin systems.} with Hamiltonian:
\begin{equation}
    H(\Pi,Q) = \f12 \sum_{i\in\L}  \Pi_i^2 +U(Q) \; , \quad\qquad U(Q)= - \sum_{i,j\in\L} J_{ij} Q_i Q_j + \sum_{i\in\L} V(Q_i) \; ,
\end{equation}
where $\L \subset \mathbb{Z}^d$ is a finite $d$-dimensional lattice, and $J_{ij}$ is not restricted to nearest neighbouring sites, but rather assumed to be of the form:
\begin{equation}
    J_{ij} \sim \f{1}{|i-j|^{d+2\z}} \;  , \quad\qquad 0<\z<1  \; ,
\end{equation}
with the case $\z=1$ being equivalent to the nearest-neighbour (or short-range) model.

In the classical canonical partition function, the configuration variables $Q$ and their conjugate momenta $\Pi$ are independent variables (e.g. taking values in $\mathbb{R}$), and expectation values in the canonical ensemble are computed as integrals over phase space with the Boltzmann weight. When considering a $Q$-dependent observable $\cA(Q)$, the integral over $\Pi$ factors out:
\begin{equation}
    \la \cA(Q) \ra_{\rm class.} \propto \int \, e^{-\b U(Q)} \cA(Q) \, \prod_{i\in\L} dQ_i \; .
\end{equation}
In the formal continuum limit $\L\to \mathbb{R}^d$ and $Q_i\to\phi(x)$ this becomes a Euclidean functional integral, and $\b$ is absorbed into the definition of the couplings of the action $\b U(Q) \to S[\phi]$:
\begin{equation} \label{eq:introLR}
    S[\phi] \sim - \int d^d x d^d y \, \f{\phi(x)\phi(y)}{|x-y|^{d+2\z}} + \int d^d x\, V(\phi) \; .
\end{equation}

In the quantum case, $\Pi$ and $Q$ are instead replaced by non-commuting operators $\hat{\Pi}$ and $\hat{Q}$, and expectation values computed as $\la \cA(\hat{Q}) \ra_{\rm quant.} \propto \Tr[\cA(Q) e^{-\b \hat{H}}]$. Noticing that $e^{-\b \hat{H}}$ has the form of the time evolution operator $e^{\imath t \hat{H}}$ with imaginary time $t=\imath \b$,  one then proceeds using the Trotter formula to split the interval $[0,\b]$ in $n$ steps, and introducing resolutions of the identity in both the $|\Pi\ra$ and $|Q\ra$ bases at each step, thus reducing again to commuting variables. Integrating over $\Pi$, one reaches:
\begin{equation} \label{eq:QModel}
    \la \cA(\hat{Q}) \ra_{\rm quant.} \propto \int \, e^{-\f{1}{2\epsilon} \sum_{<k,l>} \sum_{i\in\L} (Q_{i,k}-Q_{i,l})^2 - \epsilon \sum_{k} U(Q_k)} \cA(Q) \, \prod_{k=1}^n \prod_{i\in\L} dQ_{i,k} \; ,
\end{equation}
where $\epsilon=\b/n$ and the lattice now includes a (Euclidean) time direction, with periodic boundary condition $Q_{i,n+1}=Q_{i,1}$, along which the action is short-range ($<k,l>$ denotes nearest-neighbours).
In the continuum limit, this becomes again a Euclidean field theory, but on a $(d+1)$-dimensional space, and with an anisotropy in the (Euclidean) time direction, if $\z\neq 1$.\footnote{
Notice that such a construction goes through in a very similar fashion for any Hamiltonian of the form $H(\Pi,Q)=K(\Pi)+U(Q)$, with the only difference that if the integral in $\Pi$ is not Gaussian then the result will not be Gaussian in $(Q_{i,k}-Q_{i,l})$; but it will still be short-ranged in time.
For a case with mixing terms, see for example the rotor model \cite{Sachdev:2011fcc}.}

The main point of this example is to highlight that given a model with long-range interactions in space, its classical and quantum partition functions do not differ just in the number of dimensions, but also in the structure of the action entering the path integral.
Notice that this does not preclude the possibility of interpreting as a finite-temperature quantum system the model \eqref{eq:introLR} with a compactified direction: in fact, for $\z\leq 1$ the model is reflection positive and the Osterwalder-Schrader theorem \cite{Osterwalder:1973dx} grants the existence of an underlying quantum system. However, the construction of the corresponding quantum Hamiltonian can be very complicated and even include additional degrees of freedom. For example, it is known that an action like \eqref{eq:introLR} with $d=1$ can be obtained from a spin-boson model, i.e.\ a single quantum spin coupled to a dissipative bosonic bath, by a similar procedure as the one leading to \eqref{eq:QModel} and by tracing out the bosonic degrees of freedom (e.g.\ \cite{Winter_2009}).

In this paper, we will be interested in distinguishing and studying the same kind of long-range model in two different regimes: as a classical statistical field theory with one compact spatial direction and as a quantum statistical field theory at finite temperature (i.e.\ with compact Euclidean time).
In other words, we will consider a continuous version of the model \eqref{eq:QModel}, defined in general on $S^1_{\beta}\times S^1_{L}\times \mathbb{R}^{d-1}$, with long-range interaction in the $d$ spatial directions and standard local interaction in the imaginary-time direction, i.e.\ the coordinate on $S^1_{\beta}$ (see equation \eqref{eq:aniso-action} below for the explicit form of the action). However, we will never consider explicitly the case with both $\b$ and $L$ finite: we will start with the classical system at $\b\to 0$ and $0<L<\infty$ (i.e.\ a purely long-range model on $S^1_{L}\times \mathbb{R}^{d-1}$), and then study the finite-temperature quantum system at $0< \b \leq \infty$ and $L\to\infty$ (i.e.\ a short-range in time and long-range in space model on $S^1_{\b}\times \mathbb{R}^{d}$).

In both cases, we will focus on the critical model. 
As usual, this will correspond to tuning the non-compactified version of the model (i.e. on $\mathbb{R}^d$ or  $\mathbb{R}^{d+1}$) to the infrared fixed point of the renormalization group, where the model is scale invariant.
In the isotropic case  (the classical model), it is known that such invariance is promoted to full conformal invariance \cite{Paulos:2015jfa}, and as we review below, an interesting question is what happens to the conformal structure of the theory once a spatial direction is compactified.
In the anisotropic case (the quantum model), the situation is more complicated, as the $SO(d+1)$ group of rotations is broken to $SO(d)$ even at zero temperature. The scaling invariance is in this case of the Lifshitz type (see below), and at present it is not clear to us whether this will be enhanced to some form of anisotropic conformal invariance (see \cite{henkel2010:vol2} for examples of that).

\paragraph{Finite-size effects in CFTs.}  

In general, a conformal field theory (CFT) in $\mathbb{R}^d$ is unambiguously defined by its CFT data, that is, the spectrum of scaling dimensions of its primary operators and the set of all the operator product expansion (OPE) coefficients. One-point functions vanish identically, two-point functions are fixed up to normalization by the spin and scaling dimensions of the operators, and three-point functions are fixed by the same data plus the OPE coefficients. For higher $n$-point functions no new data are needed, as in principle they can be reconstructed by repeated use of the OPE.  Things change when one or more spatial directions are compactified. In fact, such compactifications obviously introduce a scale (the size $L$ of the compact direction) that breaks both scale and rotation invariance.
Luckily, not all the beauty of CFT is lost: if the system is still tuned to the same critical point corresponding to the CFT of the non-compact case, we expect that the compactification will not affect small-distance\footnote{Smaller than the smallest size of the compact directions.} properties, such as the OPE and the short-scale divergence of the two-point functions (hence the scaling dimensions). 
As a consequence, the reduction of $n$-point functions to lower correlators is still possible, at least for configurations of the $n$ points inside a ball of radius smaller than $L$. 
On the other hand, this reasoning does not apply to one-point functions: since translations are preserved by toroidal compactifications, one-point functions have to be constant, and thus do not admit a short-distance limit.
Therefore, as a consequence of the broken scale invariance, the one-point function of an operator $\cO$ can acquire a non-vanishing value, proportional to $L^{-\D_\cO}$, where $\D_\cO$ is the scaling dimension of the operator.
The non-vanishing coefficients of one-point functions thus become new CFT data, that permeate through OPE to all the higher $n$-point functions.

Such aspects of finite-size/temperature effects in CFTs have been discussed in \cite{Petkou:1998fb,Petkou:1998fc,Iliesiu:2018fao,Petkou:2018ynm,David:2023uya,Diatlyk:2023msc}, and we follow in particular the presentation of \cite{Iliesiu:2018fao}.
On $S^1_{L}\times\mathbb{R}^{d-1}$, conformal symmetry constrains the one-point functions to take the form: 
\begin{equation} \label{1ptFunctionCFT}
    \langle \mathcal{O}^{\mu_1\cdots\mu_J} \rangle=\frac{b_{\mathcal{O}}}{L^{\Delta_{\mathcal{O}}}}\left( e^{\mu_1} \cdots e^{\mu_J}-\text{traces}\right) \; ,
\end{equation}
where $\Delta_{\mathcal{O}}$ and $J$ are the scaling dimension and spin (order of the symmetric traceless tensor representation of $SO(d)$) of the operator $\mathcal{O}$, and $e$ is the unit vector in the compact direction. The $L$ dependence in \eqref{1ptFunctionCFT} is fixed but the coefficient $b_\mathcal{O}$ is not: it is the new CFT data we discussed above. The non-vanishing of $b_\mathcal{O}$ affects the OPE of the two-point functions, which otherwise would only include the identity operator. Indeed, starting with a two-point function in direct space:
\begin{equation}
    G(y,\mbx)=\langle\phi(y, \mbx)\phi(0,0)\rangle \; , \qquad y\in S^1_{L}\; , \;  \mbx\in \mathbb{R}^{d-1} \; ,
\end{equation}
and using the OPE for the two fields $\phi$ (which is convergent in the region $x^2=y^2+\mbx^2<L^2$, as argued in \cite{Iliesiu:2018fao}) one obtains: 
\begin{equation}\label{eq.2ptfunctionOPE}
    G(y,\mbx)=\sum_{\mathcal{O}\in\phi\times\phi} \frac{f_{\phi \phi \mathcal{O}}}{c_{\mathcal{O}}} |x|^{\Delta_{\mathcal{O}}-2\Delta_{\phi}-J}x_{\mu_1}\cdots x_{\mu_J} \langle \mathcal{O}^{\mu_1\cdots\mu_J}(0) \rangle \; ,
\end{equation}
where $f_{\phi \phi \mathcal{O}}$ is the OPE coefficient and $c_{\mathcal{O}}$ is the normalization coefficient of the two-point function of $\cO$.
As the tensor structure in \eqref{1ptFunctionCFT} is symmetric and traceless one can use \cite{Dolan:2011dv,Costa:2016hju}:
\begin{equation} \label{eq.ContractionSymTraceless}
    |x|^{-J}x_{\mu_1}\cdots x_{\mu_J} \left( e^{\mu_1}\cdots e^{\mu_J}-\text{traces} \right)=C^{\left(\frac{d-2}{2}\right)}_J(\eta) \frac{J! \Gamma(\frac{d-2}{2})}{2^J \Gamma(\frac{d-2}{2}+J)} \; ,
\end{equation}
where $\eta=y/|x|$ and $C^{\left(\frac{d-2}{2}\right)}_J(\eta)$ is a Gegenbauer Polynomial and rewrite the OPE of the two-point function as:
\begin{equation} \label{Conf2ptOPE}
G(y,\mbx)=\sum_{\mathcal{O}\in \phi \times \phi} \frac{a_{\mathcal{O}}}{L^{\Delta_{\mathcal{O}}}} C^{\left(\frac{d-2}{2}\right)}_J(\eta) |x|^{\Delta_{\mathcal{O}}-2\Delta_{\phi}} \, , \qquad
  a_{\mathcal{O}}\equiv \frac{b_{\mathcal{O}}f_{\phi \phi \mathcal{O}}}{c_{\mathcal{O}}} \frac{J! \Gamma(\frac{d-2}{2})}{2^J \Gamma(\frac{d-2}{2}+J)} \;.
\end{equation}

Note that in the infinite-size case one can go through the same steps and write an OPE expansion for the two-point function, but since only
the term corresponding to the identity operator (with $\D_{\cO}=0$) has a non-vanishing expectation value in the limit $L\rightarrow\infty$ one recovers the standard conformal form of the two-point function.

In the finite-size case one can in principle read off the coefficients $a_{\mathcal{O}}$ by expanding the two-point function as a sum over terms of the form  $C^{\left(\frac{d-2}{2}\right)}_J(\eta) |x|^{\Delta-2\Delta_\phi}$. Conversely, once the $a_{\mathcal{O}}$ coefficients are known one can reconstruct the two-point function. 
In practice, both of these reconstructions are typically hard, as they require knowing exactly either the two-point function or the CFT data.
However, they can become useful in special circumstances in which either of these pieces of information is available, as in the large-$N$ limit that we will employ here.

The new data $b_{\cO}$, for $\cO$ in the OPE of two fundamental fields of the short-range $O(N)$ model, have been computed in \cite{Iliesiu:2018fao,Diatlyk:2023msc} at leading and subleading order of the $1/N$ expansion. Here we will tackle similar computations for the long-range case.

\paragraph{Fractional Lifshitz field theories.}
As argued above, the quantum version of a long-range model, at zero or non-zero temperature, is mapped to a Euclidean field theory with a strong anisotropy. That is, it is not just the compactification of the Euclidean time direction that breaks rotation invariance: the invariance is broken already  by the local dynamics. The model that arises in such case (see our action \eqref{eq:aniso-action}, as well as \cite{Dutta:2001,Defenu:2017}) is a kind of Lifshitz field theory, that is, the type of theory used as Landau free energy of a Lifshitz point \cite{chaikin2000principles}.
Lifshitz points are characterized by a distinctive phase diagram, including a modulated phase, and an anisotropic scale invariance:
\begin{equation} \label{eq:LifshitzScaling}
    \tau \to \Om^z \tau, \quad x_i\to \Om x_i \; ,
\end{equation}
where we denoted $\tau$ the Euclidean time and the exponent $z$ is known as \emph{dynamic critical exponent}, because of the similarity with the scale invariance of dynamic critical phenomena \cite{Hohenberg:1977ym}.
Typical Lifshitz field theories are scalar field theories whose kinetic term has two time derivatives\footnote{In the classical interpretation of Lifshitz points, ``time" is actually one spatial direction of a spatially anisotropic classical system \cite{chaikin2000principles}. However, an interpretation as time has also appeared in the context of quantum Lifshitz points \cite{Ardonne:2003wa} or of Lorentz-violating quantum field theories \cite{Anselmi:2007ri}.} and $2z$ spatial derivatives, with $z$ typically being an integer (most commonly $z=2$).
Our action will be of a similar form, but with non-integer $z=\z$, fixed by the corresponding classical long-range model \eqref{eq:introLR}, and thus we will call it \emph{fractional Lifshitz field theory} (FLFT).

Given that most of the literature on Lifshitz field theories is focused on actions with $z=2$, or other integer values, there are several open questions for our FLFT, even at zero temperature. For example, as we mentioned above, we do not currently know whether at the interacting fixed point scale invariance is enhanced to some form of anisotropic conformal invariance. Moreover, we do not know yet if at such fixed point the dynamic critical exponent takes a value different from $\z$,\footnote{As a reminder, in the isotropic long-range model there is no anomalous dimension.} and we also do not know if above some $\z^*<1$ there is a crossover to the short-range universality class ($\z=1$), as in the isotropic long-range model. We hope to address these questions in a separate paper (the last two require in particular to go beyond the leading order of the large-$N$ approximation, to which we stick in the present work), while here we wish to use this model for a comparative study of some finite-temperature versus finite-size effects.

\paragraph{Plan of the paper.} The paper is organised as follows. In section \ref{sec:GFFT}, we start by studying the case of a generalized free field theory on $S_L^1\times \mathbb{R}^{d-1}$, as a warm up.  In the process we need to deal with higher twist operators that are absent in the short-range case. We  reproduce the results of \cite{Iliesiu:2018fao} for the 
 $a_{\cal O}$ coefficients in the expansion of the two-point function for operators with arbitrary spin and twist by a different technique.
In section \ref{sec:finitesize}, we study the long-range $O(N)$ model with finite size. We prove the existence of a finite-size mass and compute the coefficients $b_{\mathcal{O}}$ for spinning operators with minimal twist. We also compute the one-point functions for operators with arbitrary twist, and deduce their $a_{\cal O}$  coefficients. In section \ref{sec:finitetemp}, we switch to the case of the quantum long-range model at finite temperature, described by a FLFT on $S^1_{\b}\times \mathbb{R}^{d}$. We show that at zero temperature this model admits an interacting infrared fixed point. We then show that a thermal mass is generated when putting the model on $S^1_{\b}\times \mathbb{R}^{d}$ and we derive the one-point functions for an infinite family of scaling operators at finite temperature. 

In appendix \ref{app:bilinear}, we give some details on the computation of the zero temperature CFT data for bilinear operators, that is the normalizations of the two-point functions and the OPE coefficients $f_{\phi \phi J}$.

\section{Generalized free field theory}
\label{sec:GFFT}

As a warm up, we first consider the case of a conformal generalized free field  theory (GFFT), i.e. a massless Gaussian theory with a non-local inverse covariance, also known as mean field theory \cite{Karateev:2018oml} or fractional Gaussian field \cite{lodhia2016fractional}. It is worth discussing it in some detail because it is the simplest case of (non-local) CFT, and because typical long-range models can be defined as perturbations of a GFFT.

We begin by recalling some basic facts in the non-compact case, viewed as a classical statistical field theory at criticality, that is, the GFFT defined on $\mathbb{R}^d$, with Euclidean signature.
By definition this is a CFT whose only non-vanishing connected $n$-point function is the two-point function, given by the covariance $C(x)$ of a Gaussian measure, which however has a non-canonical scaling exponent $\D_{\phi}> d/2-1$:
\begin{equation} \label{eq:freeC}
C(x) =\frac{c(\Delta_{\phi})}{|x|^{2\Delta_{\phi}}} \; , \qquad c(\Delta_{\phi})=\frac{\Gamma(\Delta_{\phi})}{2^{d-2\Delta_{\phi}}\pi^{d/2}\Gamma(\frac{d}{2}-\Delta_{\phi})} \; .
\end{equation}
Writing $\D_{\phi}=d/2-\z$, such GFFT can be obtained from a functional integral with the action:\footnote{We typically assume $0<\z<1$. The restriction to $\z<1$ is imposed to preserve reflection positivity (unitarity in Lorentzian signature), but also because $\z>1$ would correspond to a strong short-range rather than long-range action, and moreover the operator with $\z=1$ would in that case be a relevant perturbation. The restriction to $\z>0$ is instead chosen to avoid a strong long-range action, with its associated unusual thermodynamic features  \cite{Campa:2009rev}.}
\be \label{eq:GFFT}
S_{\rm GFFT}[\phi] = \f12 \int \dd x  \, \phi(x) (-\partial^2)^{\z} \phi(x) \; ,
\ee
where $\partial^2=\p_\m\p^\m$ is the Laplace operator.
The fractional power of the Laplacian can be defined in many equivalent ways \cite{Kwasnicki:2017}, and in particular as the inverse of the operator displayed in \eqref{eq:freeC}, which has a kernel of the same form, but with the replacement $\D_{\phi}\to d-\D_{\phi}$, leading to the kinetic term of \eqref{eq:introLR}.\footnote{Notice that $c(d-\D_{\phi})=c(d/2+\z)<0$ for $0<\z<1$, thus matching the minus sign in  \eqref{eq:introLR}. Notice also that despite such minus sign, the GFFT action is positive definite, as best understood in momentum space: this is because the position space representation of the fractional Laplacian is singular and needs a subtraction, which is not written explicitly in \eqref{eq:introLR} (see footnote 18 of \cite{Benedetti:2021wzt}).}
The easiest definition is of course in Fourier space, where $(-\p^2)^{\z}$ is defined as the multiplication operator 
$(p^2)^\z$, which is the inverse of the Fourier transform of \eqref{eq:freeC} (the coefficient $c(\D_{\phi})$ is chosen precisely to have a unit coefficient in momentum space, $\tilde{C}(p) =1/(p^2)^\z$).

Since the theory is Gaussian, the scaling dimensions of local composite operators are simply given by $n_{\phi}\D_{\phi}+n_{\p}$, where $n_{\phi}$ is the number of $\phi$ fields and $n_{\p}$ the number of derivatives.
This is the same as in the local free field theory, but the operator content is slightly different in the two cases: in the GFFT case, objects like the energy-momentum tensor or  the field equations are non-local, hence they do not appear in the algebra of local operators.
As a consequence, we cannot use the Schwinger-Dyson equation:
\begin{equation}
    \la \f{\d S}{\d \phi(x_0)} \phi(x_1) \cdots \phi(x_n) \ra = \sum_{i=1}^n \d(x_0-x_i) \la \prod_{j\neq i}^{1,\ldots,n} \phi(x_j) \ra \;,
\end{equation}
to conclude that insertions of $(-\partial^2)\phi$ in any $n$-point function (at separate points)  vanish.\footnote{In fact it is straightforward to check that they do not vanish: using the fact that the theory is Gaussian we reduce $\la [\p^2 \phi](x_0) \phi(x_1) \cdots \phi(x_n) \ra$ to products of two-point functions, and we observe that $\la [\p^2 \phi](x_0) \phi(x_i)\ra= \p^2_{x_0}\la \phi(x_0) \phi(x_i)\ra \neq \d(x_0-x_i)$ for $\D_{\phi}\neq d/2-1$.} 

Another consequence of Gaussianity is that in the OPE of two fundamental fields, $\phi\times \phi$, only bilinear operators in $\phi$ appear, that is, operators of the schematic form:
\begin{equation} \label{BilinearPrimariesgfft}
     [\phi\phi]_{k,\m_1 \cdots \m_J}(x)=  \, :\phi(x) (\partial^2)^k \left( \partial_{\mu_1}...\partial_{\mu_J}-\text{traces} \right)\phi(x) : \; ,
\end{equation}
where $J\in 2\mathbb{N}_0$, $k\in\mathbb{N}_0$, and the colon notation implies a renormalization of the composite operator (normal ordering in this case).
These operators have dimension:
\begin{equation}
    \D_{k,J}= 2\D_{\phi} + J + 2k \; .
\end{equation}
Introducing the twist $\t_{k,J}=\D_{k,J}-J$, we see that the operators with $k=0$ are the minimal-twist operators, also known as double-twist operators \cite{Komargodski:2012ek}.
Once more, the appearance of the higher-twist operators, with $k>0$, marks a difference with respect to the local free theory.
Another way to see this difference is by using the conformal partial wave expansion \cite{Dobrev:1975ru,Simmons_Duffin_2018,Liu:2018jhs,Gurau:2019qag} of the four-point function. A conformal four-point function of identical scalars can be written as follows:
\begin{equation}
    \begin{split}
        \langle \phi(x_1)\phi(x_2)\phi(x_3)\phi(x_4) \rangle=&\langle \phi(x_1)\phi(x_2)\rangle \langle \phi(x_3)\phi(x_4) \rangle+\\
        & \ \ + \sum_{J} \int_{\frac{d}{2}-\imath \infty}^{\frac{d}{2}+\imath\infty}\frac{dh}{2\pi \imath} \ \frac{1}{1-k(h,J)} \, \mu^d_{\Delta_{\phi},J} (h,J) \, G^{\Delta_{\phi}}_{h,J}(x_i) \; ,
    \end{split}
\end{equation}
where $G^{\Delta_{\phi}}_{h,J}(x_i)$ is the conformal block, $\mu^d_{\Delta_{\phi},J} (h,J)$ is a known function\footnote{$\mu^d_{\Delta_J} (h,J)$ is the Plancherel weight $\rho(h,J)$ multiplied by the right normalization of three-point functions, necessary to make them an orthonormal basis of the appropriate space of bilocal functions. } and $k(h,J)$ is the eigenvalue of the Bethe-Salpeter four-point kernel. In the case of a free theory, we have $k(h,J)=0$ and one can close the contour of integration of $h$ on the right of the complex plane and pick the poles of $\mu^d_{\Delta_{\phi},J} (h,J) \, G^{\Delta_{\phi}}_{h,J}(x_i)$ on the right of $\frac{d}{2}$. The location of such poles are the scaling dimensions of bilinear primaries and the residues are their OPE coefficients with two $\phi$. In \cite{Benedetti:2019ikb} it is shown that for $\D_{\phi} > \D_{SR}=\frac{d-2}{2}$ the location of poles are at $h=\Delta_{k,J}=2\Delta_{\phi}+J+2k$, but in the limit $\Delta_{\phi}\rightarrow\D_{SR}$ all the residues with $k\ge1$ go to zero, hence those poles disappear.

\bigskip

As written in \eqref{BilinearPrimariesgfft}, the bilinear operators are in a schematic form because we have not taken into account the primary constraint, ensuring that they transform appropriately under the full conformal group.
In the short-range case the precise form of bilinear primaries has been extensively studied \cite{Dobrev:1975ru,Craigie:1983fb,Skvortsov:2015pea} and we review it in appendix \ref{app:bilinear}. However, as we already emphasized above, in that case $k$ is forced to be zero by the equation of motion. 
As a consequence, the operators with $k>0$, relevant for the GFFT, have not been studied as much in the literature, and to the best of our knowledge their general form is not known, although a recurrence relation was derived in \cite{Fitzpatrick:2011dm}. 
Below we give the explicit expressions for some primary operators with $k=1$ and $k=2$:
\begin{equation}\label{PrimariesK1and2}
    \begin{split}
        &\frac{[\phi\phi]_{1,0}(x)}{\mathcal{N}_{1,0}} = :\phi \partial^2 \phi(x): +\left(\frac{\Delta_{SR}}{\Delta_{\phi}}-1\right) :\partial_\alpha \phi \partial_\alpha \phi (x): \; , \\
        &\frac{[\phi\phi]_{2,0}(x)}{\mathcal{N}_{2,0}}=:\phi \partial^2 \partial^2 \phi(x):+ \left( 4\frac{ \left( \Delta_{SR}-\Delta_{\phi}-1\right)}{\Delta_{\phi} }\right):\partial_\alpha\phi \partial_\alpha \partial^2 \phi(x):\\ &\ \ \ \ \ \ \ \ \ \ \ \ \  \ \ +\left( \frac{\left(\Delta_{\phi} -\Delta _{SR}+1\right) \left(\Delta_{\phi}  (\Delta_{\phi} +4)+2 -\Delta _{SR}\right)}{\Delta_{\phi}  (\Delta_{\phi} +1) \left(\Delta_{\phi} -\Delta _{SR}\right)} \right):\partial^2\phi\partial^2\phi(x): \\ &\ \ \ \ \ \ \ \ \ \ \ \ \ \ \ +\left(\frac{2 \left(\Delta_{\phi} -\Delta _{SR}\right) \left(\Delta_{\phi} -\Delta _{SR}+1\right)}{\Delta_{\phi}  (\Delta_{\phi} +1)}\right) : \partial_\alpha \partial_\beta \phi \partial_\alpha \partial_\beta \phi(x): \; ,\\
        &\frac{[\phi\phi]_{1,\m_1\m_2}(x)\xi^{\m_1}\xi^{\m_2}}{\mathcal{N}_{1,2}}=:\phi\partial^2 (\imath \xi \cdot \partial)^2 \phi(x):+\left( \frac{(\Delta_\phi +2) (\Delta_\phi +3)}{\Delta_\phi  (\Delta_\phi +1)} \right):\partial^2\phi (\imath \xi \cdot \partial)^2 \phi(x): \\
        & \ \ \ \ \ \ \ \ \ \ \ \ \ \ \ +\left(\frac{2 \left(\Delta _{\text{SR} }-\Delta_\phi\right)}{\Delta_{\phi} }\right):\partial_\alpha \phi (\imath \xi \cdot \partial)^2 \partial_\alpha \phi(x):-\left(\frac{2 (\Delta_\phi +3)}{\Delta_\phi }\right) :\phi \partial^2 (\imath \xi \cdot \partial) \phi (\imath \xi \cdot \partial) \phi(x): \\
        & \ \ \ \ \ \ \ \ \ \ \ \ \ \ \ \ +\left( \frac{2 (\Delta_\phi +2) \left(\Delta_\phi -\Delta_{\text{SR}}\right)}{\Delta_\phi  (\Delta_\phi +1)} \right):\partial_\alpha(\imath \xi \cdot \partial)\phi \partial_\alpha (\imath \xi \cdot \partial) \phi(x): \; .
    \end{split}
\end{equation}
where we recall that $\Delta_{SR}=\frac{d-2}{2}$ is the scaling dimension of the field in the short-range case, $\xi$ is a complex vector with $\xi^2=0$ and $\mathcal{N}_{k,J}$ is a normalization that we are free to fix at convenience.

We notice that for $\D_{\phi}=\D_{SR}$ only terms with at least one $\p^2\phi$ survive, but, as explained above, such terms vanish in the short-range case due to the field equations. Also, we notice that the operators with $k>0$ do not coincide with the trace part subtracted in the spinning operators. For example, from the formulas of appendix \ref{app:bilinear} we have:
\begin{equation}
\begin{split}
    \frac{[\phi\phi]_{0,\m\n}(x)}{\mathcal{N}_{0,2}} = & :\phi \p_\m\p_\n \phi(x): -\frac{\Delta_{\phi}+1}{\Delta_{\phi}} :\p_\m \phi \p_\n \phi (x):  - \f{\d_{\m\n}}{d} \left( :\phi \partial^2 \phi(x): -\frac{\Delta_{\phi}+1}{\Delta_{\phi}}  :\partial_\alpha \phi \partial_\alpha \phi (x): \right)\; ,
\end{split}
\end{equation}
and we see that the trace part is a different linear combination than the one appearing in $[\phi\phi]_{1,0}(x)$.

\subsection{Propagator on $S_L^1\times \mathbb{R}^{d-1}$}

As anticipated in the introduction, we now place the model in a hyper-strip geometry with one compactified dimension of length $L$ and periodic boundary conditions.
The propagator in momentum space is just:
\be
\tilde{C}_L(q_n,\mbp) = \frac{1}{(\mbp^{2}+q_n^2)^\zeta} \; ,
\ee
with $q_n = 2\pi n/L $, $n\in\mathbb{Z}$, and $\mbp\in\mathbb{R}^{d-1}$. The subscript $L$ signifies that the covariance corresponds to a compactified strip geometry, and in the non-compact case we recover the GFFT covariance, $C_{\infty} \equiv C$.
In order to Fourier transform back to direct space, we will introduce a fictitious integration in $q$ via $1=\int d q \, \delta (q-q_n)$ and use the Poisson summation formula for distributions:
\begin{equation} \label{eq:Poisson}
    \frac{1}{L}\sum_{n} \delta(q-q_n)=\frac{1}{2\pi}\sum_{m} e^{\imath L m q} \;.
\end{equation}

Considering $q$ as the first component of a $d$-dimensional continuous momentum $p=(q,\mbp)$, we get:
\begin{equation} \label{eq:Fourier-freeG}
    C_L(y,\mbx)= \frac{1}{L} \sum_{n=-\infty}^{+\infty} \int \frac{d^{d-1} \mbp}{(2\pi)^{d-1}} \frac{e^{\imath \mbp \cdot \mbx + \imath q_n y}}{(\mbp^2+q_n^2)^\zeta} =\sum_{m=-\infty}^{+\infty} \int \frac{d^d p}{(2\pi)^d} \frac{e^{\imath p\cdot x_{m}}}{(p^2)^{\zeta}} \; ,
\end{equation}
where $y\in S_L^1$, $\mbx\in\mathbb{R}^{d-1}$, and $x_m=(y+m L, \mbx)$.
At this stage we could proceed by using a Schwinger parametrization for the denominator (as for example in \cite{Benedetti:2020rrq}). However this does not generalize nicely to the massive case of the next section, hence we evaluate here the two-point function in position space by a different method.

We write the momentum space propagator as a Stieltjes transform (or K\"all\'en-Lehmann representation):
\begin{equation} \label{eq:C-Fourier}
  \f{1}{(p^2)^\z} = \int_0^{+\infty} ds\, \f{\r(s)}{p^2+s} \;,\qquad
\r(s) = \frac{\sin(\pi \z)}{\pi} s^{-\z} \; .
\end{equation}
The Fourier transform in \eqref{eq:Fourier-freeG} is then reduced to the standard one for the short-range massive scalar free propagator, and we thus obtain:
\begin{equation} \label{eq:freeG}
\begin{split}
    C_L(y,\mbx) & =\frac{\sin(\pi \z)}{(2\pi)^{d/2}\pi} \sum_{m=-\infty}^{+\infty} \int_0^{+\infty} ds\, s^{-\z} \, \left( \frac{\sqrt{s}}{|x_m|} \right)^{d/2-1} K_{\frac{d}{2}-1}(\sqrt{s} |x_m|) \\
    & = \frac{\sin(\pi \z)}{(2\pi)^{d/2}\pi} \sum_{m=-\infty}^{+\infty} \frac{1}{|x_m|^{d-2\z}}  \int_0^{+\infty} ds\, s^{-\z+d/4-1/2} \,  K_{\frac{d}{2}-1}(\sqrt{s}) \\
    & = \frac{\sin(\pi \z) \Gamma(1-\z)\Gamma(d/2-\z)}{2^{2\z}\pi^{d/2}\pi} \sum_{m=-\infty}^{+\infty} \frac{1}{|x_m|^{d-2\z}} \\
    &= c(\D_\phi) \sum_{m=-\infty}^{+\infty} \frac{1}{|x_m|^{2\D_\phi}} \; ,
\end{split}
\end{equation}
where $ K_{\frac{d}{2}-1}(x)$ is the modified Bessel function of second kind. 
This might look like a useless exercise, as we knew already the result of the Fourier transform in \eqref{eq:Fourier-freeG}, but its usefulness lies in highlighting a strategy that will be fruitful in the massive case.

\subsection{One-point functions of the minimal-twist operators}

We are now interested in computing the one-point function of bilinear primary operators introduced in \eqref{BilinearPrimariesgfft}.
The most convenient way to deal with the tracelessness condition of higher-spin operators is by means of the index-free formalism of \cite{Dobrev:1975ru,Craigie:1983fb}: define an operator $\mathcal{O}_{k,J}(x,\xi)$ by contracting each derivative with a complex null vector $\xi= (\chi,\bm{\xi})$ such that $\bm{\xi}^2=-\chi^2$ (see appendix \ref{app:bilinear} for more details). For now, we will consider $k=0$. 

When computing in momentum space the one-point function of the primary operator defined in \eqref{eq:spin-Op} we just have to substitute $\pleft \rightarrow -\imath p$ and $\pright\rightarrow\imath p $. As a result, only the $n=0$ term in the sum \eqref{eq:f-Gegenbauer} survives, the prefactor simplifies to one and we get:
\begin{equation}
\begin{split}
    \langle\mathcal{O}_{0,J}(x,\xi)\rangle &\equiv \xi^{\m_1} \cdots \xi^{\m_J} \langle [\phi_a\phi_a]_{0,\m_1 \cdots \m_J}(x) \rangle \\
    &=\frac{1}{L} \sum_{n=-\infty}^{+\infty} \int \frac{d^{d-1} \mbp}{(2\pi)^{d-1}} \frac{(\imath \mbp \cdot \bm{\xi}+ \imath q_n \chi)^J}{(\mbp^2+q_n^2)^\zeta} - \int \frac{d^d p}{(2\pi)^d} \frac{(\imath p\cdot \xi)^J}{(p^2)^{\zeta}}\; ,
\end{split}
\end{equation}
where the subtraction term comes from the normal ordering of the operator, as implemented before compactifying one direction, and an ultraviolet regularization is implicitly assumed. 
As written, it is not obvious that the normal ordering is sufficient in making the one-point function finite when we remove the regularization. In order to see that indeed it is finite, it is convenient to use the Poisson formula, and rewrite:
\begin{equation}
\begin{split}
    \langle\mathcal{O}_{0,J}(x,\xi)\rangle  &= \sum_{m=-\infty}^{+\infty} \int \frac{d^{d} p}{(2\pi)^{d}} \frac{(\imath \mbp \cdot \bm{\xi}+ \imath q \chi)^J \, e^{\imath L m q }}{(\mbp^2+q^2)^\zeta} - \int \frac{d^d p}{(2\pi)^d} \frac{(\imath p\cdot \xi)^J}{(p^2)^{\zeta}} \\
    & = \sum_{m\neq 0} \int \frac{d^{d} p}{(2\pi)^{d}} \frac{(\imath \mbp \cdot \bm{\xi}+ \imath q \chi)^J \, e^{\imath L m q }}{(\mbp^2+q^2)^\zeta}\; .
\end{split}
\end{equation}
The oscillating factor renders each term in the sum finite (as can be shown by deforming the contour of integration of $q$ in the complex plane), and the normal ordering removes the only term without oscillating factor.
Actually, the subtracted term vanishes for $J>0$ in any regularization preserving rotation invariance, such as a cutoff on $p^2$ (via the insertion in the integrand of a cutoff function $f(p^2/\L^2)$ decaying sufficiently fast at infinity): by symmetry reasons the resulting integral over $p$ must be proportional to a linear combination of products of $J/2$ Kronecker delta functions, which vanish when contracted with the null vector $\xi$.
The $m\neq 0$ terms are instead not rotation invariant, due to the phase factor depending only on one momentum component, and thus are non-vanishing.

Following an idea from \cite{Diatlyk:2023msc}, we define a generating function for these expectations:
\begin{equation} \label{eq:generFunct-GFFT}
    \sum_{J\ge0}\frac{ \langle\mathcal{O}_{0,J}(x,\xi)\rangle}{J!} =\sum_{m\neq 0} \int \frac{d^d p}{(2\pi)^d} \frac{e^{\imath p\cdot \xi_{m}}}{(p^2)^{\zeta}} \equiv C^{(0)}(\xi) \; ,
\end{equation}
where $\xi_{m}=(\chi+m L,\bm{\xi})$.
We recognize that such a generating function is the two-point function, with the $m=0$ term subtracted, and evaluated at the null vector $\xi$, $C^{(0)}(\xi) =C_L (\xi)- C(\xi)$. Therefore, it is enough to use our previous results, evaluated at $x=\xi$  and read off the coefficients of the one-point function from the expansion in powers of  $\chi$.

Note that the general OPE formula for the two-point function in \eqref{Conf2ptOPE}, when evaluated at a null $\xi$ (and subtracting the divergent contribution from the identity operator, $\D_\cO=J=0$), becomes:
\begin{equation}
    C^{(0)}(\xi)= \sum_{J\in2\mathbb{N}_0}\frac{1}{L^{2\D_\phi+J}} \frac{b^{(\text{free})}_{0,J} f_{\phi\phi J}}{c_{0,J}} \chi^J \; .
\end{equation}
On the other hand, using the general formula \eqref{1ptFunctionCFT}, and contracting with our null vector $\xi$, we find:
\begin{equation} \label{eq:xi-contraction}
    \langle\mathcal{O}_{0,J}(x,\xi)\rangle     = \f{b^{(\text{free})}_{0,J} }{L^{2\D_\phi +J}}
    \chi^J \;.
\end{equation}
Therefore, by comparison to \eqref{eq:generFunct-GFFT}, in our conventions  it must be $f_{\phi \phi J }/c_{0,J}=1/J!$. In appendix \ref{app:bilinear} we prove that this is indeed the case.

After subtracting the divergent $m=0$ term (corresponding to the contribution from the identity operator) from \eqref{eq:freeG}, and replacing $x\to\xi$, we have:
\begin{equation}
    C^{(0)}(\xi)=c(\Delta_\phi)\sum_{m\neq 0} \frac{1}{|L m |^{2\Delta_\phi}(1+\frac{2\chi}{m L})^{\Delta_\phi}} \; .
\end{equation}
Expanding in $\chi$, and combining the terms with $m>0$ and $m<0$, the odd powers of $\chi$ cancel while the even ones reconstitute a Riemann zeta function, which we denote  $\zeta_R$, to avoid confusion with our long-range parameter:
\begin{equation}
        C^{(0)}(\xi) =c(\Delta_\phi) \sum_{J \in 2\mathbb{N}_0} \frac{\chi^J}{L^{J+2\Delta_\phi}} \frac{\Gamma(J+\Delta_\phi)}{ J!\Gamma(\Delta_\phi)} 2^{J+1}  \zeta_R(J+2\Delta_\phi) \; .
\end{equation}
From such expansion we can extract the one-point function coefficients:
\begin{equation}
    b^{(\text{free})}_{0,J}=c(\Delta_\phi)\frac{\Gamma(J+\Delta_\phi)}{ \Gamma(\Delta_\phi)} 2^{J+1}  \zeta_R(J+2\Delta_\phi)  \; .
\end{equation}

Lastly, using the relation $f_{\phi \phi J }/c_{0,J}=1/J!$ we find:
\begin{equation}
    \begin{split}
        a^{(\text{free})}_{0,J}&\equiv \frac{J! \Gamma(\D_{SR})}{2^J \Gamma(\D_{SR}+J)} \frac{b^{(\text{free})}_{0,J} f_{\phi \phi J }}{c_{0,J}}=c(\Delta_\phi)\frac{2  \Gamma(\Delta_{\phi} +J)\Gamma(\D_{SR}) }{\Gamma(\D_{\phi})\Gamma(\D_{SR} +J) }\zeta_R (J+2\Delta_\phi ) \; ,
    \end{split}
\end{equation}
which matches with the result in \cite{Iliesiu:2018fao}, up to the different normalization of the two-point function $c(\Delta_\phi)$.

\subsection{One-point functions of higher-twist operators}\label{sec.1ptHigherTwistsFree}

As we already discussed, in the case of long-range models the OPE of $\phi\times\phi$ contains also higher-twist operators, of the form \eqref{BilinearPrimariesgfft} with $k>0$. 
As we do not have a general formula for the primaries with $k>0$, we cannot explicitly compute their two-point functions, and the associated coefficient, or their OPE coefficient. However, the precise distribution of derivatives is irrelevant for the tadpole computation, as the derivatives acting on the left and right field $\phi$ in Fourier space both reduce to (plus or minus) the same loop momentum.
Therefore, we can implicitly define the normalization to be the one such that the one-point function reduces to:\footnote{In the cases of the operators that we wrote in  \eqref{PrimariesK1and2}, the explicit normalizations are 
\[
    \mathcal{N}_{1,0}=\frac{\Delta_{\phi}}{\Delta_{SR}} \, , \quad \mathcal{N}_{2,0}= \frac{\Delta_{\phi}  (\Delta_{\phi} +1) \left(\Delta_{\phi} -\Delta _{SR}\right)}{2\Delta_{\phi}\left(\Delta_{SR}^2+\Delta_{SR}+1\right)-\Delta_{SR}^2\left(2\Delta_{SR}+1\right)+\Delta_{SR}+2} \, , \quad \mathcal{N}_{1,2}=-\frac{\Delta_{\phi}\left(\Delta_{\phi}+1\right)}{2\Delta_{SR}} \, .
\]
}
\begin{equation}
    \langle\mathcal{O}_{k,J}(x,\xi)\rangle = \frac{1}{L} \sum_{n=-\infty}^{+\infty} \int \frac{d^{d-1}\mbp}{(2\pi)^{d-1}} \frac{(\imath \mbp \cdot \bm{\xi}+ \imath q_n \chi)^J}{(\mbp^2+q_n^2)^\z} (\mbp^2+q_n^2)^k  
    -\int \frac{d^{d}p}{(2\pi)^{d}} \frac{(\imath x \cdot \xi)^J}{(p^2)^\z} (p^2)^k\; ,
\end{equation}
which is finite, by the same argument used for $k=0$, and again the subtraction is really needed only for $J=0$.
As before, we will aim at a generating function for the corresponding coefficient $a_{k,J}$, defined as in \eqref{Conf2ptOPE}, which is independent of the normalization of $\cO_{k,J}$.

We begin by defining a generating function of such one-point functions. That is, after using again the Poisson formula, we write:
\begin{equation} \label{eq:C(k)Definition}
\begin{split}
    C^{(k)}(\chi,\bm{\xi}) &\equiv\sum_{J\ge0}\frac{ \langle\mathcal{O}_{k,J}(x,\xi)\rangle}{J!} =\sum_{m\neq 0} \int \frac{d^d p}{(2\pi)^d} \frac{e^{\imath p\cdot \xi_{m}}}{(p^2)^{\zeta}} (p^2)^k \\
    &= (-\p^2)^k C^{(0)}(x)_{|_{x=(\chi,\bm{\xi})}}
    \;,
\end{split}
\end{equation}
which unsurprisingly is the result of applying $k$ Laplacians to the (subtracted) covariance and evaluating it at $x=\xi$.
Then we can find an alternative representation of $C^{(k)}(\chi,\bm{\xi})$ just by acting with $k$ Laplacians on \eqref{eq.2ptfunctionOPE} and evaluating at $x=\xi$, we find:
\begin{equation}\label{eq.C(k)Expansion}
    C^{(k)}(\chi,\bm{\xi})=(-1)^k \sum_{J\in 2\mathbb{N}_0} \frac{\chi^J}{J!} \frac{a_{k,J}}{L^{2\Delta_\phi+2k+J}} \frac{2^{J+2k}k!\Gamma(k+J+\D_{SR}+1)}{(J+\D_{SR})\Gamma(\D_{SR})} \; .
\end{equation}
On the other hand, from the Fourier representation in \eqref{eq:C(k)Definition} it is clear that  $C^{(k)}(\chi,\bm{\xi})$ corresponds to the replacement $\zeta\rightarrow\zeta-k$ (i.e.\  $\Delta_\phi \rightarrow\Delta_\phi+k$) in the subtracted covariance $C^{(0)}(\chi,\bm{\xi})$, and thus we obtain:
\begin{equation}
    C^{(k)}(\chi,\bm{\xi})  = c(\D_\phi+k) \sum_{m\neq 0} \frac{1}{|\xi_m|^{2\D_\phi+2k}} \; ,
\end{equation}
which, after repeating the same steps as before, becomes:
\begin{equation}
    \begin{split}  C^{(k)}(\chi,\bm{\xi})&=c(\Delta_\phi+k) \sum_{J \in 2\mathbb{N}_0} \frac{\chi^J}{L^{J+2k+2\Delta_\phi}} \frac{\Gamma(J+\Delta_\phi+k)}{ J!\Gamma(\Delta_\phi+k)} 2^{J+1}  \zeta_R(J+2\Delta_\phi+2k) \; .
    \end{split}
\end{equation}
Lastly, by comparison with \eqref{eq.C(k)Expansion}, we extract the $a_{k,J}$ coefficients:
\begin{equation}
    \begin{split}
        a_{k,J}
        &=c(\Delta_\phi)  \frac{2 (J+\D_{SR} ) \Gamma(\D_{SR})\Gamma(\D_{\phi}+J+k)\Gamma(\D_{\phi}-\D_{SR}+k)}{k! \Gamma(\D_{\phi})\Gamma(\D_{SR}+J+k+1)\Gamma(\D_{\phi}-\D_{SR})} \zeta_R (J+2 k+2 \Delta_\phi ) \; ,
    \end{split}
\end{equation}
that again match with the result in \cite{Iliesiu:2018fao} up to the $c(\Delta_\phi)$ factor.

\section{Classical long-range $O(N)$ model with a finite size}
\label{sec:finitesize}

The long-range $O(N)$ model is defined by the action:
\begin{equation} \label{eq:iso-action}
    S[\phi]=\int_{\mathbb{R}^d} d^dx \left(\frac{1}{2}\phi_a(x) (-\p^2 )^\z \phi_a(x)+\frac{1}{2} M_{\text{bare}}^{2\z} \phi^2(x)+\frac{\l}{4 N}(\phi^2(x))^2\right) \; ,
\end{equation}
where $a=1,\ldots,N$, $\phi^2(x)\equiv\sum_a \phi_a(x)^2$, and $M_{\text{bare}}$ is the bare mass, which we treat as a pure counterterm because we are interested in the critical theory.
It is convenient to introduce also the intermediate field representation, by the usual Hubbard-Stratonovich transformation:
\begin{equation} \label{eq:ON-action-mixed}
    S[\phi,\s]=\int d^dx \left(\frac{1}{2}\phi_a(x) (-\p^2 )^\z \phi_a(x)+\frac{1}{2} M_{\text{bare}}^{2\z} \phi^2(x) 
    + N \f{\s(x)^2}{4\l} + \f{\imath}{2} \, \s(x)  \phi^2(x) \right) \; .
\end{equation}

For any $\z  \in(d/4,1)$ one finds a non-trivial IR fixed point, most easily accessible in the large-$N$ expansion, or in the small-$\epsilon$ expansion where $\z=(d+\epsilon)/4$ \cite{Fisher:1972zz}. 
Characteristic features of such fixed point are the absence of anomalous dimension (that is, the two-point function scales as in the corresponding GFFT, as proved rigorously in \cite{Lohmann:2017}), and $\z$-dependent critical exponents. Such $\z$-dependent fixed points are then associated to a one-parameter family of universality classes, collectively called \emph{long-range universality class}.
An interesting crossover phenomenon actually occurs before reaching $\z=1$: there exists a $\z^\star\in(d/4,1)$ such that for $\z>\z^\star$ the operator $\phi_a\p^2\phi_a$ (i.e.\ the local kinetic term) becomes relevant at the IR fixed point, thus destabilizing it and driving the system to the short-range universality class \cite{Sak:1973}.  The value of such crossover exponent for $2\leq d<4$ is $\z^\star=1-\eta_{SR}/2$, where $\eta_{SR}$ is the anomalous dimension of the short-range model.\footnote{For $d=1$ the upper limit is $\z^\star= 1/2$, beyond which there can be no spontaneous symmetry breaking \cite{Dyson:1968up}.} However, since  $\eta_{SR}=O(1/N)$ \cite{Wilson:1972cf}, at large $N$ there is no crossover, and our analysis will apply all the way to $\z=1$.

\begin{figure}[htbp]
    \centering
    \includegraphics{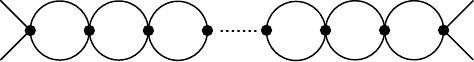}
    \caption{Chain of bubbles with $n\geq 1$ vertices. The connected four-point function is given by the sum of all such chains.}
    \label{fig:bubblechain}
\end{figure}

In the large-$N$ limit, the propagator only receives mass corrections, via tadpoles, and we thus have the following Schwinger-Dyson equation:
\begin{equation} \label{eq:SD-LO-noncomp}
    \tilde{G}(p)^{-1}=\tilde{C}(p)^{-1}+M_{\text{bare}}^{2\z}+\frac{\lambda}{2} \int \frac{d^{d} p'}{(2\pi)^{d}} \tilde{G}(p') \; ,
\end{equation}
where $\tilde{C}(p)$ is as in the GFFT, and we defined $\la\phi_a(x) \phi_a(0)\ra = \d_{ab} G(x)$ and $\tilde{G}(p)$ the Fourier transform of $G(x)$.
Since the tadpole is momentum-independent, the solution must be of the form $\tilde{G}(p)^{-1}=\tilde{C}(p)^{-1}+M^{2\z}$.
Tuning the model to criticality requires choosing a bare mass $M_{\text{bare}}$ such that $M=0$, which amounts to canceling the tadpole contribution, hence we have:
\begin{equation} \label{eq:bare_mass}
    M_{\text{bare}}^{2\z} = - \frac{\lambda}{2} \int \frac{d^{d} p}{(2\pi)^{d}} \f{1}{(p^2)^{\z}} \equiv -\frac{\lambda}{2} \cT \; .
\end{equation}
In the same limit, the connected four-point function is given by a sum of chains of bubble diagrams, with full leading-order propagator on each edge, see Figure~\ref{fig:bubblechain}. 

Therefore, the proper vertex in momentum space reads:
\be
\G^{(4)}(p) =  \f{2\l}{1+\l B(p)}\; ,
\ee
where $p$ is the total transferred momentum and we defined the bubble integral:
\begin{equation}
    B(p) = \int \f{d^d q}{(2\pi)^d} \f{1}{q^{2\zeta} (p-q)^{2\zeta}} =  \f{c(d/2-\z)^2}{c(d-2\z)} |p|^{d-4\z} \; .
\end{equation}

We define the renormalized dimensionless coupling $g$ at some renormalization group scale $|p|=\m$:
\be
g=  \m^{4\D_{\phi}-d} \G^{(4)}(\m) /2 \; .
\ee
It follows that the beta function is:\footnote{Notice that for $d-4\D_{\phi}=4\z-d=\eps$ (i.e.\ for $\z=(d+\eps)/4$) the second term is finite (and positive) in the limit $\eps\to0$, because $c(2\D_{\phi})\sim1/\G(\eps/2)$, and we recover the standard one-loop result $\b(g)=-\epsilon g +  2 g^2 /((4\pi)^{d/2}\Gamma(d/2))$ (see \cite{Benedetti:2020rrq}).} 
\be
\b(g)= \m\f{d g}{d\m} = -(d-4\D_{\phi}) g +  \f{c(\D_{\phi})^2}{ c(2\D_{\phi})} (d-4\D_{\phi}) g^2 \; .
\ee
Therefore, for $\D_\phi<d/4$ (i.e.\ $\z>d/4$) the beta function has a zero (i.e.\ a fixed point) at $g^{\star}=  c(2\D_{\phi})/c(\D_{\phi})^2$.
It is worth noticing that since we can trivially invert the relation between bare and renormalized coupling to get:
\be
\m^{4\D_{\phi}-d} \l = \f{g}{1-\f{c(\D_{\phi})^2}{c(2\D_{\phi})} g}\; ,
\ee
the fixed point corresponds to the limit $\l\to\infty$.

Another well known large-$N$ result for the short-range $O(N)$ model, which generalizes straightforwardly to the long-range case, is the fact that at the IR fixed point the composite operator $\phi^2(x)$ (of dimension $\D_{0,0} = 2\D_{\phi}=d-2\z$ in the GFFT) is mapped to its shadow operator (of dimension $\D_{\s} = d-\D_{0,0} =2\z $), which can be identified with the intermediate field $\s(x)$. This is a particular realization of the general result by Gubser and Klebanov \cite{Gubser:2002vv} on large-$N$ flows driven by a double-trace operator (in this case, the $(\phi^2)^2$ interaction). There are many other ways to see the same result, for example by means of the conformal partial wave representation of the four-point function \cite{Benedetti:2021wzt}, or diagrammatically, as we will see below.

The other bilinear primaries \eqref{BilinearPrimariesgfft} remain instead in the spectrum.
In addition, in the interacting model we also have other operators, with more powers of $\s$ or $\phi$.
Naively, at large $N$ such operators are suppressed in the OPE of $\phi\times\phi$.
However, the situation is more subtle: as argued in \cite{Iliesiu:2018fao}, in the expansion \eqref{Conf2ptOPE} it can happen that for certain $\cO$'s while $f_{\phi\phi\cO}/\sqrt{c_{\cO}}$ is suppressed, $b_\cO/\sqrt{c_{\cO}}$ is boosted, thus resulting in a finite contribution to $a_\cO$ even at large $N$. This in fact does happen for the operators $\sigma^n(x)$ which yield non-trivial contributions to the OPE, as we will discuss in \ref{sec:higher-twist_int}.

\paragraph{Two-point functions of spinning operators.}
When evaluating at leading order in $1/N$ the two-point functions of bilinear operators of the $O(N)$ model, we need to sum chains of bubble diagrams, as for the four-point function. We now show that for the spinning operators at the fixed point the sum over the chain of bubbles reduces to the contribution of a single bubble, and thus their two-point functions (and hence scaling dimensions) are the same as in the GFFT.

First, we check once again that the $\phi^2$ operator drops out of the CFT.
The sum of bubbles becomes a simple geometric sum in Fourier space, hence we have:
\begin{equation}
    \la [\phi_a\phi_a]_{0}(x_1) [\phi_a\phi_a]_{0}(x_2) \ra
    = \int \f{d^d p}{(2\pi)^d} e^{\imath p\cdot x_{12}} \f{B(p)}{1+\l B(p)} \; .
\end{equation}
For $\l\to\infty$ the result vanishes as expected.
The two-point function of $\s(x)$ is similar, but with the replacement $B(p)\to \l$ in the numerator (it coincides with the $\G^{(4)}(p)$ above), hence it remains finite in the limit. We infer that $\phi^2$ is replaced by its shadow operator $\s$ at the interacting fixed point.

For higher-spin operators we have instead to distinguish the first and last bubbles:
\begin{equation}
    \la [\phi_a\phi_a]_{0,\m_1 \cdots \m_J}(x_1) [\phi_a\phi_a]_{0,\n_1 \cdots \n_J}(x_2) \ra
    = \int \f{d^d p}{(2\pi)^d}  e^{\imath p\cdot x_{12}}
    \left( B_{\m_1 \cdots \m_J,\n_1 \cdots \n_J} - B_{\m_1 \cdots \m_J} \f{\l}{1+\l B(p)} B_{\n_1 \cdots \n_J}  \right) \; ,
\end{equation}
where $B_{\m_1 \cdots \m_J,\n_1 \cdots \n_J}$ is the Fourier transform of the same two-point function in the free theory, i.e.\ the diagram without internal vertices,
while $ B_{\m_1 \cdots \m_J}$ is the Fourier transform of the two-point function of the spin-$J$ operator with the spin-0 one, also in the free theory. However, the latter vanishes: technically because of orthogonality of the Gegenbauer polynomials \cite{Craigie:1983fb}, and conceptually because in a CFT the two-point functions of primaries of different dimensions must vanish.

The explicit computation of these two-point functions is reported in appendix \ref{app:bilinear}.

\subsection{Schwinger-Dyson equation in the hyper-strip geometry}

We now place the model in a hyper-strip geometry with one finite dimension of length $L$ and periodic boundary conditions. The action is as in \eqref{eq:iso-action} with the replacement $\mathbb{R}^d\to S^1_L \times \mathbb{R}^{d-1}$.
Such setting has been extensively studied in the short-range case, and in particular in the large-$N$ limit (e.g.\ \cite{Moshe:2003xn,Sachdev:1993pr,Petkou:1998fb,Petkou:1998fc,Iliesiu:2018fao,Petkou:2018ynm,Diatlyk:2023msc}).

Like in the short-range model, also in the long-range case the propagator will acquire a mass due to the introduction of the compact direction, as we now show  by reducing the Schwinger-Dyson equation to a mass gap equation.

At large $N$, the Schwinger-Dyson equation in momentum space is:
\begin{equation} \label{eq:SD-LO}
    \tilde{G}(q_{n_0}, \mbp_0)^{-1}=\tilde{C}_L(q_{n_0},\mbp_0)^{-1}+M_{\text{bare}}^{2\z}+\frac{\lambda}{2} \frac{1}{L}\sum_n\int \frac{d^{d-1} \mbp}{(2\pi)^{d-1}} \tilde{G}(q_{n},\mbp) \; .
\end{equation}
The inverse free propagator is $\tilde{C}_L(q_{n},\mbp)^{-1}=(\mbp^2+q_{n}^2)^\z$ as in the GFFT, $M_{\text{bare}}^{2\z}$ is the bare mass \eqref{eq:bare_mass}, and $q_n = 2\pi n/L$, as before. 
As the tadpole term in \eqref{eq:SD-LO} does not depend on the external momentum, the full inverse two-point function is $\tilde{G}(q_{n},\mbp_0)^{-1}=(\mbp_0^2+q_{n}^2)^\z+M_L^{2\z}$ for some $M_L$  to be determined self-consistently.

The self-consistent gap equation for the finite-size mass is:
\begin{equation}
\begin{split}
M_L^{2 \zeta} &= M_{\text{bare}}^{2 \zeta}+ \frac{\lambda}{2} \frac{1}{L} \sum_{n} \int \frac{d^{d-1} \mbp}{(2 \pi)^{d-1}} \frac{1}{ (\mbp^2+q_n^2)^\zeta+M_L^{2 \zeta}} \\
&= \frac{\lambda}{2}\left( - M_L^{2\zeta}\int \frac{d^{d} p}{(2 \pi)^{d}} \frac{1 }{((p^{2})^{\zeta}+M_L^{2\zeta})(p^{2})^{\zeta}} +\sum_{m \ne 0}  \int \frac{d^{d} p}{(2 \pi)^{d}}\frac{e^{\imath L m q}}{(p^2)^{\zeta}+M_L^{2\zeta}} \right) \\
& \equiv \frac{\lambda}{2}\, \left( \cT(M_L) -\cT \right)
\; ,
\end{split}
\end{equation}
where in the second equality we used the Poisson summation formula \eqref{eq:Poisson}, we have introduced again a $d$-dimensional momentum $p=(q,\mbp)$, and we have combined the $m=0$ term with the explicit expression of the bare mass. In the last step we have defined the subtracted tadpole $ \cT(M_L) -\cT$.

The first integral in the subtracted tadpole can be computed explicitly and we find:
\begin{equation} \label{eq:T_subtracted}
        \cT(M_L) -\cT = \frac{\pi  M_L^{d-2 \zeta }}{(4 \pi)^{d/2} \G(d/2) \zeta \sin \left(\frac{d \pi }{2 \zeta }\right)}+\sum_{m \ne 0}  \int \frac{d^{d} p}{(2 \pi)^{d}}\frac{e^{\imath L m  q}}{(p^2)^{\zeta}+M_L^{2\zeta}} \; .
\end{equation}

The second term is not manifestly UV convergent, but its convergence can be inferred from the convergence of the $M_L=0$ case of the previous section (see \eqref{eq:freeG} without the $m=0$ term, evaluated at $x_m=(L m, \mathbf{0})$, giving $b^{(\text{free})}_{0,0}$), as the inclusion of the mass does not make the UV behavior worse.\footnote{We can also explicitly show its convergence deforming the contour of integration of $q$ to a Hankel contour in the upper (lower) complex plane for positive (negative) $m$.}
Its explicit evaluation is however rather complicated, and we will take a different route to the gap equation below.

As we recalled above, at the IR fixed point the bare coupling $\l$ flows to infinity, hence the mass gap equation corresponds to finding the value of $M_L$ such that the subtracted tadpole \eqref{eq:T_subtracted} vanishes.
We emphasize that the subtraction is essential to the existence of a solution, as otherwise the tadpole would seem positive, which of course is naive as it is in fact divergent: for $d/4<\z<1$ and $2<d<4$ (and for $1/4<\z<\z^\star=1/2$ if $d=1$) the sine in the first term is negative, and thus a cancellation between the two terms can occur.

We will come back to this equation
below, from a different perspective, and prove that it admits a solution.

\subsection{Long-range two-point function with mass}

The two-point function in direct space reads:
\begin{equation} \label{FourierTransf2pt}
    G(y,\mbx)= \frac{1}{L} \sum_{n=-\infty}^{+\infty} \int \frac{d^{d-1} \mbp}{(2 \pi)^{d-1}} \frac{e^{\imath \mbp\cdot \mbx+\imath y q_n}}{(\mbp^2+q_n^2)^{\zeta}+M_L^{2\zeta}} \;.
\end{equation}
As before, we introduce $1= \int d q \delta (q-q_n)$ and we use the Poisson summation formula to rewrite it as:
\begin{equation}
    G(y,\mbx)=\sum_{m=-\infty}^{+\infty}\int \frac{d^d p}{(2\pi)^d} \frac{e^{\imath p \cdot x_{m}}}{(p^2)^{\zeta}+M_L^{2\zeta}}
    \;,
\end{equation}
where $p$ is a $d$-dimensional momentum and $x_m=(y+m L, \mbx)$. 

In order to proceed further, we rewrite again the momentum-space propagator $\tilde{G}(p) = \left((p^2)^{\zeta}+M_L^{2\zeta}\right)^{-1}$ as a Stieltjes transform. While such a rewriting was trivial in the massless case because the spectral density had to be scaling invariant, it is slightly more involved in the presence of a mass.
First, as the momentum-space propagator is only a function of the squared momentum, we can define $g(p^2)= \tilde{G}(p)$, and we use Cauchy's theorem to write:
\begin{equation}
g(p^2) = \f{1}{2\pi\imath} \oint_{\g} \f{g(z)}{z-p^2} dz \;,
\end{equation}
with $\g$ a counterclockwise contour encircling $z=p^2$. Then, we deform the contour to infinity, and in the process we pick up poles and branch cuts of $g(z)$.
Now, for $\z\in(0,1)$, the function $g(z)=1/(z^\z+M_L^{2\z})$ has a branch cut on the negative axis, and no poles in the principal sheet.\footnote{The poles are at $z=M_L^2 e^{\f{\imath \pi (1+2k)}{\z}}$, for integer $k$, and it can be seen that for $0<\z<1$ none of them is in the principal sheet, which by definition corresponds to ${\rm Arg}(z)\in(-\pi,\pi)$.}
Therefore, the deformation of the contour $\g$ leads to a Hankel contour around the cut, which in turn leads to the standard formula for the inversion of the Stieltjes transform
$\r(s) = \f{1}{2\pi\imath} \left( g(-s-\imath\epsilon) - g(-s+\imath\epsilon)  \right) $, and we find:
\begin{equation} \label{SpDensity}
    \tilde{G}(p)  =  \f{1}{(p^2)^{\z}+M_L^{2\z}} =   \int_0^{+\infty} ds\, \f{\r(s)}{p^2+s} \;, \qquad
    \r(s) = \frac{\sin(\pi \z)}{\pi} \frac{s^\z}{s^{2\zeta}+2 M_L^{2\zeta} s^{\zeta} \cos(\pi \zeta)+M_L^{4\zeta}} \; .
\end{equation}
Going back to position space, and setting $s=t^2$, we have:
\begin{equation} \label{eq:G_KL}
    G(y,\mbx)=\frac{2 \sin(\pi \z)}{(2\pi)^{d/2} \pi}\sum_{m=-\infty}^{+\infty} \int_{0}^{+\infty} d t \; \left(\frac{t}{|x_m|} \right)^{d/2-1} K_{\frac{d}{2}-1}(t |x_m|)  \frac{t^{1+2\zeta}}{t^{4\zeta}+2 M_L^{2\zeta} t^{2\zeta} \cos(\pi \zeta)+M_L^{4\zeta}} \; .
\end{equation}
Because of the mass, in this case we cannot factor out of the integral the $x$-dependence.

At this step one might worry that for $\zeta\to 1$ everything vanishes because of the $\sin(\pi \z)$ factor. 
This is not the case because in this limit the denominator in the integrand becomes $(t^{2}-M_L^{2})^2$ and the integral develops a singularity (this is the analogue of the $\G(1-\z)$ factor in the massless case of \eqref{eq:freeG}).
In fact in the $\zeta \to 1$ limit we recover the short-range case \eqref{eq:freeG},
as it can be shown by using a variation of the Plemelj--Sokhotski formula.

For later reference we write explicitly the case $d=3$, in which the Bessel function takes a simple form:
\begin{equation} \label{Gconvergent}
   G(y,x)=\frac{\sin(\pi \z)}{2\pi^2}\sum_{m=-\infty}^{+\infty} \int_{0}^{+\infty} d t \; \frac{e^{-t |x_m|}}{|x_m|}  \frac{t^{1+2\zeta}}{t^{4\zeta}+2 M_L^{2\zeta} t^{2\zeta} \cos(\pi \zeta)+M_L^{4\zeta}} \; .
\end{equation}

\subsection{Finite-size mass solution}

\begin{figure}[htbp]
\centering
\includegraphics[scale=0.45]
{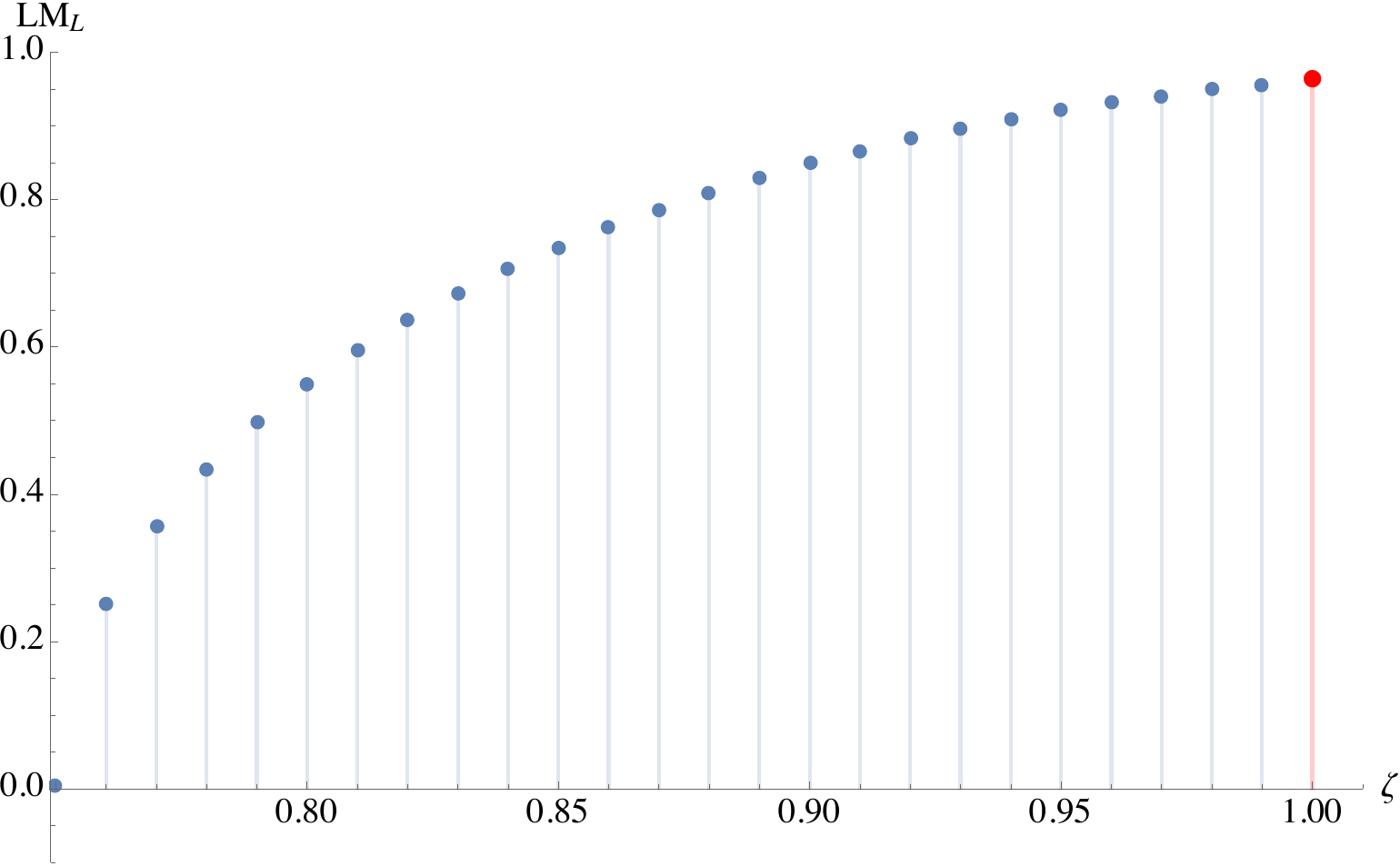}
\caption{Numerical solutions for the mass gap equation as a function of $\z$, at $d=3$. The red dot is the value of the short-range thermal mass and it is approached in the $\z\rightarrow1$ limit as expected.}
\label{fig:TheMass2}
\end{figure}

The $m\neq 0$ part of the subtracted tadpole \eqref{eq:T_subtracted} 
is the $m\neq 0$ part of the two-point function, evaluated at $x=0$.
Therefore, we can use the representation \eqref{Gconvergent} to rewrite the  mass gap equation in $d=3$ as:
\begin{equation}
\frac{ L M_L}{4 \pi  \zeta \sin  \left(\frac{3 \pi }{2 \zeta }\right)}-\frac{\sin(\pi \z)}{\pi^2}  \int_{0}^{+\infty} d t \;  \frac{t^{2 \zeta +1}  }{t^{4 \zeta }+2 \cos (\pi  \zeta ) t^{2 \zeta }+1} \ln \left(1-e^{-L M_L t}\right)=0 \;,
\end{equation}
which is more suitable for a numerical evaluation, having eliminated the sum in favor of a single integral that is manifestly convergent. 
We find that the equation always admits a real positive solution, which we plot as a function of $\z$ in Figure~\ref{fig:TheMass2}. The solution goes to zero at $\z=3/4$, where the IR fixed point merges with the Gaussian one, and it goes to the short-range value (e.g.\ \cite{Sachdev:1993pr}) for $\z\to 1$.

\subsection{One-point functions of the minimal-twist operators}
\label{subsec:spinningL}

The one-point functions of the minimal-twist operators in the hyper-strip geometry are defined similarly to the GFFT case, except for the presence of the mass and the restriction to $J>0$, as the $\phi^2$ operator is not in the spectrum:
\begin{equation}
\begin{split}
    \langle\mathcal{O}_{0,J}(x,\xi)\rangle  
    & = \sum_{m\neq 0} \int \frac{d^{d} p}{(2\pi)^{d}} \frac{(\imath \mbp \cdot \bm{\xi}+ \imath q \chi)^J \, e^{\imath L m q }}{(\mbp^2+q^2)^\zeta+M_L^{2\z}}\; .
\end{split}
\end{equation}
Notice that the restriction to $m\neq 0$ comes from the fact that the $m=0$ term vanishes in any rotationally invariant regularization, due to the contraction with $\xi$, and as before the other terms do not need any subtraction.

As in the previous section, following \cite{Diatlyk:2023msc}, we can evaluate the subtracted two-point function at a null vector $\xi=(\chi,\bm{\xi})$ and interpret it as the generating function of the one-point functions:
\begin{equation} \label{eq:G0_M}
    G^{(0)}(\chi,\bm{\xi})= \sum_{J> 0}\frac{ \langle\mathcal{O}_{0,J}(x,\xi)\rangle}{J!} 
    =\sum_{m\neq 0} \int \frac{d^d p}{(2\pi)^d} \frac{e^{\imath p\cdot \xi_{m}}}{(p^2)^{\zeta}+M_L^{2\z}} - M_L^{2\zeta}\int \frac{d^{d} p}{(2 \pi)^{d}} \frac{1 }{((p^{2})^{\zeta}+M_L^{2\zeta})(p^{2})^{\zeta}}\; .
\end{equation}
Notice that in order to get an exponential in the last step we had to include a spurious $J=0$ term, which however we cancel with the last term, thanks to the mass gap equation imposing the vanishing of the subtracted tadpole \eqref{eq:T_subtracted}.

An alternative explanation for the $J=0$ subtraction in \eqref{eq:G0_M} is obtained by considering the general expansion of the two-point function in \eqref{Conf2ptOPE}. As we know, at large $N$ the spectrum of the $\phi\times\phi$ OPE contains the identity operator (with $\D_{\rm id}=0$), the intermediate field (with $\D_\s=2\z$), and the bilinear operators with $\D_{k,J}=2\D_\phi+J+2k$, the $J=k=0$ operator being excluded.\footnote{As discussed earlier in the text (and originally pointed out in \cite{Iliesiu:2018fao}), one-point functions can have a positive scaling in $N$ that boosts operators whose OPE is suppressed in $1/N$, thus introducing new operators in \eqref{Conf2ptOPE} even at large $N$. However, such operators have larger scaling dimension (at fixed spin) than those we are considering here, so they do not affect this argument. \label{foot:otherOp}}
When setting $x=\xi$, since $|\xi|=0$ and the Gegenbauer polynomial is of order $J$, all the bilinear operators with $k>0$ drop from the expansion, and so does also $\s$, since $2\z>2\D_\phi$ for $\z>d/4$.
Therefore, we should find only the $\D_{0,J>0}$ contributions plus a divergent one from the identity operator. The latter corresponds to the massless propagator in the non-compact case, but the $m=0$ term we have removed in the first term of \eqref{eq:G0_M} is the massive propagator. In order to get the right operator content of \eqref{Conf2ptOPE}, we should thus add back this massive propagator, and then subtract the massless one, which is precisely what the last term in  \eqref{eq:G0_M} does, since:
\begin{equation}
    \int \frac{d^{d} p}{(2 \pi)^{d}} \frac{e^{\imath p\cdot x} }{((p^{2})^{\zeta}+M_L^{2\zeta}} - \int \frac{d^{d} p}{(2 \pi)^{d}} \frac{e^{\imath p\cdot x}}{(p^{2})^{\zeta}}
    =  - M_L^{2\zeta}\int \frac{d^{d} p}{(2 \pi)^{d}} \frac{e^{\imath p\cdot x}}{((p^{2})^{\zeta}+M_L^{2\zeta})(p^{2})^{\zeta}} \; ,
\end{equation}
is a function of $|x|$, convergent at $|x|=0$, and thus its value at $x=\xi$ is the same as at $x=0$.

\

\begin{figure}[htbp]
\centering
\includegraphics[scale=0.65]{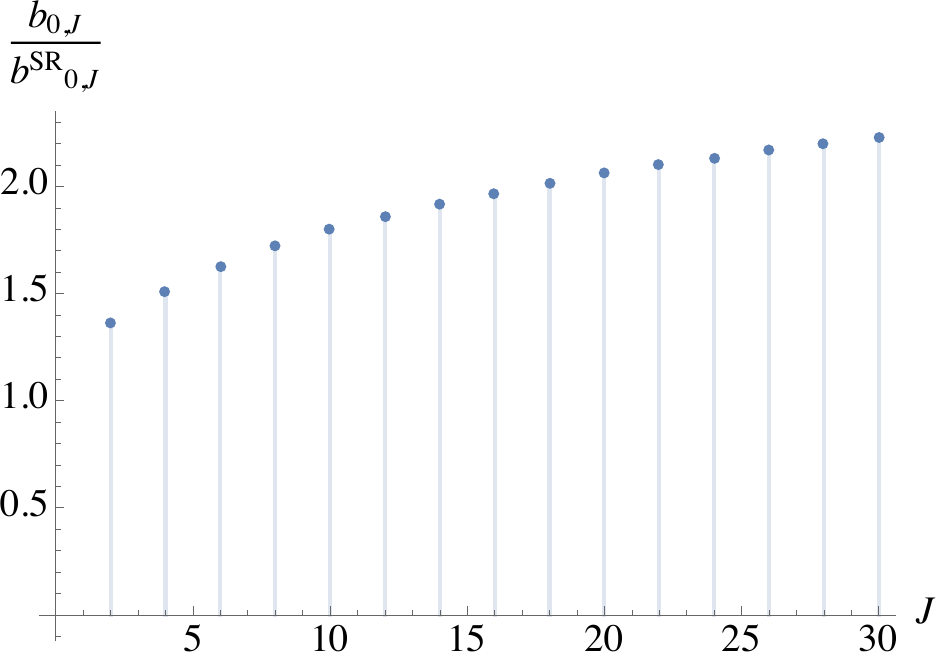}
\caption{Plot of the ratio between $b_{0,J}$ for $\zeta=0.8$ and the corresponding short-range value.
}
\label{fig:b}
\end{figure}

Subsequently we can extract the $b_{k,J}$  coefficients from the formal series expansion in $\chi$. 
In $d=3$, we have:
\begin{equation} \label{Gconvergent^0}
   G^{(0)}(\chi,\bm{\xi})=\frac{\pi  M_L^{3-2 \zeta }}{4 \pi \zeta \sin \left(\frac{3 \pi }{2 \zeta }\right)} + \frac{\sin(\pi \z)}{2\pi^2}\sum_{m\neq 0} \int_{0}^{+\infty} d t \; \frac{e^{-t |\xi_m|}}{|\xi_m|}  \frac{t^{1+2\zeta}}{t^{4\zeta}+2 M_L^{2\zeta} t^{2\zeta} \cos(\pi \zeta)+M_L^{4\zeta}} \; ,
\end{equation}
and the expansion in $\chi$ inside the integral takes the same form as in \cite{Diatlyk:2023msc}, but with $t$ replacing the mass of the short-range case. We thus get an integral representation for the one-point functions of the $J>0$ spinning operators:\footnote{Again we can use $f_{\phi \phi J }/c_{0,J}=1/J!$ and easily compute $a_{0,J}$ from $b_{0,J}$}
\begin{equation}\label{bLongRange}
    b_{0,J}=\frac{\sin(\pi \z)}{2 \pi^2} \sum_{n=0}^{J} \frac{2^{n-J+1}(2J-n)!}{n!(J-n)!}\int_{0}^{+\infty} dt \ \text{Li}_{J+1-n}\left(e^{-t L M_L}\right) \frac{t^{1+2\zeta+n}(L M_L)^{n+2-2\zeta}}{t^{4\zeta}+2 t^{2\zeta} \cos(\pi \zeta)+1} \; .
\end{equation}
Unfortunately the integral above can not be evaluated in terms of usual special functions and  needs to be evaluated numerically. We compare results to the short-range case by means of a plot in Figure~\ref{fig:b}. 

\subsection{One-point functions of higher-twist operators}
\label{sec:higher-twist_int}

As in the GFFT, we define the one-point functions of the higher-twist bilinear operators with an implicit choice of normalization such that:
\begin{equation}
    \langle\mathcal{O}_{k,J}(x,\xi)\rangle = \frac{1}{L} \sum_{n} \int \frac{d^{d-1}\mbp}{(2\pi)^{d-1}} \frac{(\imath \mbp \cdot \bm{\xi}+ \imath q_n \chi)^J}{(\mbp^2+q_n^2)^\z+M_L^{2\z}} (\mbp^2+q_n^2)^k  
    -\text{counterterms}\; .
\end{equation}
In order to make explicit the structure of divergences and related counterterms we make use again of the Poisson formula, and we write:
\begin{equation}
    \langle\mathcal{O}_{k,J}(x,\xi)\rangle  
     =  \sum_{m\neq 0} \int \frac{d^{d} p}{(2\pi)^{d}} \frac{(\imath p \cdot \xi)^J \, (p^2)^k\, e^{\imath L m q }}{(p^2)^\zeta+M_L^{2\z}}  +\d_{J,0}  \cC_k \; ,
\end{equation}
where:
\begin{equation} \label{def:Ck}
 \cC_k \equiv \int \frac{d^{d} p}{(2\pi)^{d}} \frac{ (p^2)^k}{(p^2)^\zeta+M_L^{2\z}} - \sum_{n=0}^{\lfloor \f{d+2k-2\z}{2\z}\rfloor} (-M_L^{2\z})^n \int \frac{d^{d} p}{(2\pi)^{d}} (p^2)^{k-(n+1)\z}\; .
\end{equation}
We have used again the fact that for $J>0$ the $m=0$ term vanishes in any rotationally invariant regularization. For $J=0$ we have subtracted instead the purely divergent parts of the $m=0$ term. Alternatively, we could drop such explicit subtractions and evaluate the $m=0$ term by analytic continuation in $k$ from a region of convergence (i.e.\ $-d/2<k<-d/2+\z$).
It is however interesting, and actually useful, to keep them explicit as we will see that they have a direct interpretation in terms of \eqref{Conf2ptOPE}.

We are going to define again a generating function for these one-point functions, as:
\begin{equation} \label{eq:G(k)Definition}
\begin{split}
    G^{(k)}(\chi,\bm{\xi}) &\equiv\sum_{J\ge0}\frac{ \langle\mathcal{O}_{k,J}(x,\xi)\rangle}{J!} = \sum_{m\neq 0} \int \frac{d^d p}{(2\pi)^d} \frac{e^{\imath p\cdot \xi_{m}}}{(p^2)^{\zeta}+M_L^{2\z}} (p^2)^k + \cC_k\\
    &= \left( (-\p^2)^k G(x) - \cL_k(x)\right)_{x=\xi} 
    \;,
\end{split}
\end{equation}
where:
\begin{equation} \label{def:Lk}
\begin{split}
    \cL_k(x) &= \sum_{n=0}^{\lfloor \f{d+2k-2\z}{2\z}\rfloor} (-M_L^{2\z})^n \int \frac{d^{d} p}{(2\pi)^{d}} (p^2)^{k-(n+1)\z} \, e^{\imath p\cdot x} \\
    &= \sum_{n=0}^{\lfloor \f{d+2k-2\z}{2\z}\rfloor}  \f{(-M_L^{2\z})^n\, c(d/2+k-(n+1)\z)}{|x|^{d+2k-2(n+1)\z}}\; .
\end{split}
\end{equation}
The interpretation of $\cL_k(x)$ is the following. When acting $k$ times with a Laplacian on \eqref{Conf2ptOPE} and setting $x=\xi$, we have a similar structure as in the GFFT from the bilinear operators and the identity operator. In addition, we have other operators that at fixed $k$ give vanishing contributions (see footnote \ref{foot:otherOp}), but we also have operators that give new divergent contributions from some $k$ on: these are operators $\cO$ with $\D_\cO-2\D_\phi<2k$ not even. For example, the intermediate field $\s$, having $J=0$ and non-integer dimension $\D_\s=2\z$, gives vanishing contribution at $k=0$, but a diverging one as soon as $k\geq 1$ if $d\geq 2$: indeed $\p^2|x|^{\D_\s-2\D_\phi}= \p^2|x|^{4\z-d} = (4\z-d)(4\z-2) |x|^{4\z-d-2}$, and $4\z-d-2<0$ for $\z<1$ and $d\geq 2$.
More in general, we see by comparing \eqref{def:Lk} with \eqref{Conf2ptOPE}, that the operators leading to singularities in $ (-\p^2)^k G(\xi)$ have dimension $\D_{\s^n}=2n\z$, and thus correspond to the composite fields $\s^n$:\footnote{It can be shown that spinning operators involving $\s$ or higher powers of $\phi^2$ are subleading in the $1/N$ expansion (remember that we compute $b_\cO$ at finite $L$, but $f_{\phi\phi\cO}$ and $c_\cO$ at infinite $L$). This explains why we only find contributions from $\s^n$, which has $a_{\s^n}\sim O(N^0)$.}
\begin{equation} \label{eq:deriv_sigma_n}
    (-\p^2)^k \f{a_{\s^n}}{L^{2n\z}} |x|^{2n\z-2\D_\phi} 
    = \f{a_{\s^n}}{L^{2n\z}} (-4)^{k}  \f{\G(n\z-\D_\phi +1) \G(d/2+n\z-\D_\phi )}{\G(n\z-\D_\phi +1 -k) \G(d/2+n\z-\D_\phi -k)} |x|^{2n\z-2\D_\phi-2k} \; .
\end{equation}
Therefore, we need to subtract these dangerous terms from $(-\p^2)^k G(x)$ before setting $x=\xi$, which is what we are doing with $\cL_k(x)$. Equivalently, we can introduce a regularization, subtract $\cL_k^{\rm reg.}(\xi)$, and then remove the regulator, which is what we are doing with the subtractions of $\cC_k$.

The above discussion has an important consequence: evaluating $(-\p^2)^k G(x)$ allows us to extract not only the $a_{k,J}$ coefficients (from the finite part at $x=\xi$, as we did in the GFFT), but also the $a_{\s^m}$ coefficients (from the divergent part).
The latter is obtained straightforwardly by comparing \eqref{def:Lk} and \eqref{eq:deriv_sigma_n}:
\begin{equation} \label{eq:a_sigma_n}
\begin{split}
    a_{\s^n} & =  (-1)^{n+k}(L M_L)^{2n\z}\, \f{c(\D_\phi+k-n\z)}{4^{k}} \f{\G(n\z-\D_\phi +1 -k) \G(d/2+n\z-\D_\phi -k)}{\G(n\z-\D_\phi +1) \G(d/2+n\z-\D_\phi )} \\
    & =  (-1)^n \f{(L M_L)^{2n\z}}{4^{(n+1)\z}\pi^{d/2}} \, \f{\pi}{\sin(\pi(\D_\phi-n\z)) \G(n\z-\D_\phi +1) \G(d/2+n\z-\D_\phi )}  \; .
\end{split}
\end{equation}
Notice that any dependence on $k$ has dropped out, and the result is valid for all $n\geq 1$.
In the limit $\z\to 1$ this yields the short-range result:
\begin{equation}
    a_{\s^n}^{SR} =  (L M_L)^{2n}\,c(\D_{SR}) \f{\G(1-\D_{SR})}{4^{n} \G(n+1) \G(n-\D_{SR}+1)} \; , 
\end{equation}
which in $d=3$ matches the result of \cite{Iliesiu:2018fao}.
However, notice that while in the short-range case in integer dimensions there are degeneracies (essentially all operators have integer canonical dimension), this is not the case for general $\z<1$, and therefore we can safely identify \eqref{eq:a_sigma_n} with the one-point function coefficient of $\s^n$.

Finally, we are ready to set $x=\xi$ in (\ref{eq:G(k)Definition}) and extract the $a_{k,J}$ coefficients. This can be easily done as we did in section
(\ref{sec.1ptHigherTwistsFree}), that is, by comparing the Taylor expansion of (\ref{eq:G(k)Definition}) in $\chi$ with equation (\ref{eq.C(k)Expansion}) and matching order by order the coefficients of the powers of $\chi$. The reason is that equation (\ref{eq.C(k)Expansion}) was derived starting from (\ref{Conf2ptOPE}) assuming that the dimension of operators appearing in the OPE of $\phi \times \phi$ are $\Delta_{k,J}=2\Delta_\phi+2k+J$, and this is also the case for the interacting large-$N$ theory up to the $\sigma^n$ operators that we subtracted explicitly in (\ref{eq:G(k)Definition}).
To this purpose it is convenient to write 
$\frac{(p^2)^k }{(p^2)^{\zeta}+M_L^{2\zeta}}$ as a Stieltjes transform:
\begin{equation} \label{SpDensity}
\f{p^{2k}}{(p^2)^{\z}+M_L^{2\z}} =   \int_0^{+\infty} ds\, \f{\r(s)}{p^2+s} \;, \qquad
    \r(s) = \frac{\sin(\pi \z)}{\pi} \frac{s^{\z+k}}{s^{2\zeta}+2 M_L^{2\zeta} s^{\zeta} \cos(\pi \zeta)+M_L^{4\zeta}} \ .
\end{equation}
Then specializing to $d=3$ we find:
\begin{equation}
    G^{(k)}(\chi,\bm{\xi})=\frac{\sin(\pi \z)}{2\pi^2}\sum_{m\neq0} \int_{0}^{+\infty} d t \; \frac{e^{-t |x_m|}}{|x_m|}  \frac{t^{1+2\zeta+2k}}{t^{4\zeta}+2 M_L^{2\zeta} t^{2\zeta} \cos(\pi \zeta)+M_L^{4\zeta}}+\mathcal{C}_k \ ,
\end{equation}
where $\mathcal{C}_k$ contributes to the $J=0$ term only, while the sum over $m$ will contribute to any $J$. After performing the small $\chi$ expansion we get:
\begin{equation}
    \begin{split}
        G^{(k)}(\chi,\bm{\xi}) =& -\frac{1}{L^{3-2\z+2k}}\left( \frac{\sin(\pi \z)}{\pi^2} \int_{0}^{+\infty}dt \ln \left(1-e^{-L M_L t}\right) \frac{t^{1+2\z+2k}(L M_L)^{2-2\z+2k}}{t^{4\z}+2t^{2\z} \cos(\pi \zeta)+1}\right)+\mathcal{C}_k \\
        &+\sum_{J>0}\frac{\chi^J}{J! L^{3-2\z+2k+J}}\left( \frac{\sin(\pi \z)}{2 \pi^2} \sum_{n=0}^{J} \frac{2^{n-J+1}(2J-n)!}{n!(J-n)!} \right. \crcr
       & \qquad \qquad \left. \times \int_{0}^{+\infty} dt \ \text{Li}_{J+1-n}\left(e^{-t L M_L}\right) \frac{t^{1+2\zeta+n+2k}(L M_L)^{n+2-2\zeta+2k}}{t^{4\zeta}+2 t^{2\zeta} \cos(\pi \zeta)+1} \right) \;,
    \end{split}
\end{equation}
where the first line corresponds to spin-zero operator and the sum to all the spinning operators.
Comparing with (\ref{eq.C(k)Expansion}) we finally find:
\begin{equation}
    a_{k,0}=  \frac{- \sin(\pi \z )}{(2k+1)! \pi^2}\int_{0}^{+\infty} dt \ln\left(1-e^{-t L M_L} \right)\frac{t^{1+2\z+2k}(L M_L)^{2-2\z+2k}}{t^{4\z}+2 t^{2\z} \cos(\pi \z) +1}  +\frac{(L M_L)^{3-2\z+2k}}{(2k+1)! 4\pi \z \sin \left(\pi\frac{  2 k+3}{2 \zeta }\right)}     \; , 
\end{equation}
\begin{equation}
\begin{split}
    a_{k,J}=& \frac{4^{-k-J}(J+\tfrac{1}{2})\sin(\pi \z)}{\pi^{3/2}\Gamma(J+k+\frac{3}{2})k!} \sum_{n=0}^{J} \frac{2^{n}(2J-n)!}{n!(J-n)!} \\ 
    & \qquad\qquad  \times \int_{0}^{+\infty} dt \ \text{Li}_{J+1-n}\left(e^{-t L M_L}\right) \frac{t^{1+2\zeta+n+2k}(L M_L)^{n+2-2\zeta+2k}}{t^{4\zeta}+2 t^{2\zeta} \cos(\pi \zeta)+1}  \; , 
\end{split}
\end{equation}
where we used the fact that $\mathcal{C}_k$ in $d=3$ evaluates to:
\begin{equation}
    \mathcal{C}_k=\frac{M_L^{3-2\z+2k}}{4\pi \z \sin \left(\pi\frac{  2 k+3}{2 \zeta }\right)}  \ .
\end{equation}
\

\section{Quantum long-range $O(N)$ model at finite temperature}
\label{sec:finitetemp}

The simplest example of an interacting fractional Lifshitz field theory (FLFT), of the type discussed in the introduction is described by the action:
\begin{equation} \label{eq:aniso-action}
\begin{split}
S[\phi]=\int d\tau\int d^{d}x \bigg( \frac{1}{2} & \phi_a(\tau, x) (-\p_\tau^2 ) \phi_a(\tau,x) 
 + \frac{1}{2}\phi_a(\tau,x) (-\p_i \p_i )^\z \phi_a(\tau,x)  \crcr
 & \qquad +\frac{1}{2} M_{\text{bare}}^{2\z} \phi^2(\tau,x)+\frac{\l}{4 N}(\phi^2(\tau,x))^2 \bigg)  \; ,
\end{split} 
\end{equation}
where $\p_\tau$ is the ordinary derivative with respect to the Euclidean time $\tau$, $(-\p_i \p_i)^\zeta$ is the fractional Laplacian for the $d$ spatial directions and, as before, $a=1,\ldots,N$, $\phi^2(\t,x)\equiv\sum_a \phi_a(\t,x)^2$, and $M_{\text{bare}}$ is the bare mass.
As argued in the introduction, we can view this model, when defined on $\mathbb{R}^{d+1}$ or $S_\b^1\times\mathbb{R}^d$, as describing the quantum long-range $O(N)$ model at zero or finite temperature, respectively. With the same perspective, a similar model has been considered in \cite{Dutta:2001,Defenu:2017}, where it has been studied at finite $N$ by other means.
Alternatively, one could view it as a classical anisotropic model with short-range interactions in one direction (compact or not) and long-range interactions in the remaining (non-compact) ones (see also \cite{Defenu:2016pwy} for a previous study of anisotropic long-range models).

We notice first of all that dimensional analysis of the quadratic part of the action implies that we have the mass dimensions $[x]=-1$, $[\t]=-\z$, and therefore $[\phi]=\D_\phi$ given by:
\begin{equation}
    2\Delta_\phi + 2 \zeta -\zeta - d  = 0 \quad \Rightarrow \quad
 \Delta_\phi = \frac{d-\zeta}{2} \;.
\end{equation}
As a consequence, the quartic interaction in \eqref{eq:aniso-action} is marginal (i.e.\ $[\l]=0$) for $4\Delta_\phi = d+\zeta $, that is, $\zeta = d/3$. 
The model we are interested in has $\z<1$ and a relevant quartic interaction, therefore we will concentrate on $d=2$ and $2/3< \zeta <1$, but for now we keep $d$ and $\zeta$ unspecified. 

The action \eqref{eq:aniso-action} is invariant under global $O(N)$ transformations with $\phi$ transforming in the fundamental representation. Moreover, it is clearly translation invariant, both in time and space, while for $\z\neq 1$ it has rotation invariance only in space.
We will be interested in the limiting cases in which it has also a scaling symmetry, namely, the ultraviolet Gaussian fixed point and an infrared interacting fixed point. Due to the anisotropic nature of the model we expect a Lifshitz type of scaling invariance, that is:
\begin{equation} \label{eq:Lifshitz_scaling}
    \tau \to\Omega^\z \tau, \qquad x\to \Omega x, 
    \qquad \phi(\tau,x) \to \phi(\Omega^\z \tau,\Omega x) = \Omega^{-\Delta_\phi}\phi(\tau,x) \; ,
\end{equation}
for any constant $\Omega>0$.
This symmetry is in general not promoted to a generalized form of conformal invariance.\footnote{In the case $\z=2$, when formulated as a dynamic critical model (or stochastic partial differential equation problem), it has been shown in \cite{Henkel:1993sg} that there is a larger symmetry, namely the Schr\"odinger group (see \cite{henkel2010:vol2} and references therein), but only for the response function, not the correlator. This result has been generalized in \cite{Henkel:1997zz} to the case $\z=2/n$, for $n\in\mathbb{N}$, but this does not include our interesting $d=2$ case with $\z\in(2/3,1)$.}
The anisotropy and lack of conformal invariance imply that even at fixed points the correlators are much less restricted than in the isotropic case. 
One-point functions are still vanishing (for non-compact spacetime), because of translation and scale invariance.
However, for the two-point function we have:
\begin{equation} \label{eq:G-LifshitzScaling}
    \la \phi(\t,x) \phi(0,0) \ra  = |x|^{-2\D_\phi} f_x\left( \f{|\t|^{1/\z}}{|x|} \right) = |\t|^{-\f{2\D_\phi}{\z}} f_\t\left( \f{|x|}{|\t|^{1/\z}} \right)\; ,
\end{equation}
where we have expressed it in two equivalent ways, and we expect the $f_x(u)$ and $f_\t(u)$ to be two functions that are regular in zero and vanishing at infinity as $u^{-2\D_\phi}$, but otherwise completely unconstrained.

Even if the fixed-point theories are not conformally invariant, we can still make use of the OPE, although in this case we expect it to be only an asymptotic expansion (see for example \cite{Shimada:2021xsv}). 
Indeed, assuming that a fixed-point Lifshitz field theory admits a basis in the space of operators consisting in scaling operators ${\cal O}_i$ with dimensions $\Delta_i$,  such that ${\cal O}_i(\Omega^\z \tau,\Omega x) = \Omega^{-\Delta_i} {\cal O}_i(\t,x) $,  the OPE expansion writes:
\begin{equation} \label{eq:aniso-OPE}
    {\cal O}_{i_1}(\t_1,x_1) {\cal O}_{i_2}(\t_2,x_2)   = \sum_{ k } f_{i_1 i_2 k }(\t_{12},x_{12}) \; {\cal O}_{k}(\t_2,x_2) \; ,
\end{equation}
where $\t_{12}=\t_1-\t_2$ and $x_{12}=x_1-x_2$, and the equality should be understood to be only in the asymptotic sense.
 Applying a scaling transformation and the OPE in either order we have:
\begin{equation}
\begin{split}
  {\cal O}_{i_1}(\Omega^\z \tau_1,\Omega x_1)  {\cal O}_{i_2}(\Omega^\z \tau_2,\Omega x_2)  & = \Omega^{-\Delta_{i_1} -\Delta_{i_2} }    {\cal O}_{i_1}(\t_1,x_1)  {\cal O}_{i_2}(\t_2,x_2)  \\
  & = \Omega^{-\Delta_{i_1} -\Delta_{i_2} }    \sum_k   f_{i_1i_2 i }(\t_{12},x_{12}) \; {\cal O}_k(\t_2,x_2) \;,  \crcr
  {\cal O}_{i_1}(\Omega^\z \tau_1,\Omega x_1)  {\cal O}_{i_2}(\Omega^\z \tau_2,\Omega x_2)   & = \sum_k  f_{i_1i_2 k }(\Omega^\z\t_{12},\Omega x_{12}) \; {\cal O}_k(\Omega^\z \tau_2,\Omega x_2) \\
  &= \sum_k f_{i_1i_2 k }(\Omega^\z\t_{12},\Omega x_{12}) \Omega^{-\Delta_k } {\cal O}_k( \t_2,x_2) \;, 
\end{split}
\end{equation}
hence the OPE coefficient is a scaling function 
$f_{i_1i_2 k }(\Omega^\z\t,\Omega x) = \Omega^{\Delta_k - \Delta_{i_1} -\Delta_{i_2}} f_{i_1i_2 k }(\t,x)$, that is, we can always write:
\begin{equation} \label{eq:aniso-OPEcoeff}
    f_{i j k }(\t,x) = |\t|^{\f{\Delta_k - \Delta_i -\Delta_j}{\z}} f_{i j k }\left( \f{|x|}{|\t|^{1/\z}} \right) \; .
\end{equation}
This will be useful in the finite-temperature case (compact time), in order to relate the two-point function to the one-point functions via the zero-temperature OPE coefficients.

We will first discuss the free theory as a warm up, and then move on to the interacting case.

\subsection{The massless free fractional Lifshitz field theory at zero temperature}

The action of the massless free FLFT at zero temperature is (we suppress the $O(N)$ indices here as the $N$ fields are decoupled in this case):
 \begin{equation} \label{eq:aniso-freeaction}
    S[\phi]=\int_{-\infty}^{+\infty} d\tau \int_{\mathbb{R}^d} d^{d}x 
    \left( \frac{1}{2}\phi(\tau, x) (-\p_0^2 ) \phi(\tau, x)+ \frac{1}{2}\phi(\tau, x) (-\p_i \p_i )^\z \phi(\tau, x)\right) \; .
\end{equation}
It is trivial to check that this action is invariant under Lifshitz scaling \eqref{eq:Lifshitz_scaling}.
This implies that the two-point function, or covariance $C(\t,x)$ of the Gaussian measure, must have the form \eqref{eq:G-LifshitzScaling}. Indeed, writing the direct space covariance as a Fourier transform:
\begin{equation} \label{eq:anisoC-Fourier}
    C(\tau,x)=   \int \frac{d^{d}p}{(2\pi)^{d}}  \int \frac{d\omega}{2\pi} \;  \frac{e^{\imath p \cdot x+ \imath \omega \tau}}{(p^2)^{\z}+\omega^2} \; ,
\end{equation}
the integral over $\omega$ can be performed easily, by deforming the contour and picking up a pole:
\begin{equation} \label{eq:aniso-freeC}
    C(\t,x) = \int \frac{d^{d}p}{(2\pi)^{d}} e^{\imath p \cdot x } \f{e^{-|\t|(p^2)^{\z/2}}}{2 (p^2)^{\z/2}} \; ,
\end{equation}
where the integral is convergent for $\z<d$. Simple rescalings reproduce \eqref{eq:G-LifshitzScaling} and, recalling that $2\Delta_\phi =d- \z$, we have the limit cases:\footnote{Keeping both $\t$ and $x$ different from zero, we can reduce to single integral representations in various ways. For instance, we can use spherical coordinates in \eqref{eq:aniso-freeC}, perform the angular integral and remain with an integral over the radial part, involving an hypergeometric function. Alternatively, and perhaps more interestingly, we can use again a spectral representation of the integrand, that is, perform a Stieltjes transform with respect to $p^{2}$ and then use the standard result for the short-range massive propagator, to obtain:
\[
C(\tau,x)= \int_{0}^{+\infty} \frac{ ds }{(2\pi)^{d/2+1} } \left(\frac{\sqrt{s}}{|x|}\right)^{\frac{d-2}{2}}K_{\frac{d-2}{2}}(\sqrt{s}|x|)  \; s^{-\frac{\zeta}{2}} 
 \operatorname{Im} \left(  e^{ \frac{ \imath  \pi}{2} \zeta}    \; e^{- |\tau| 
s^{\frac{\zeta}{2}}  e^{- \frac{\imath \pi}{2} \zeta}
}  \right) \; .
\]
}
\begin{equation} \label{eq:C-Lifshitz-cases}
\begin{split}
    & C(\t,0) = |\t|^{-\frac{2\Delta_\phi}{\zeta} } \int \frac{d^{d}p}{(2\pi)^{d}} \f{e^{-(p^2)^{\z/2}}}{2 (p^2)^{\z/2}}=\frac{\Gamma \left(\frac{d-\z}{\zeta }\right)}{(4 \pi )^{d/2} \zeta  \Gamma \left(\frac{d}{2}\right)} |\tau|^{-\frac{2\Delta_\phi}{\z}} \; , \\ 
    & C(0,x) = \int \frac{d^{d}p}{(2\pi)^{d}} e^{\imath p \cdot x } \f{1}{2 (p^2)^{\z/2}} = \frac{c(\D_{\phi})}{2} |x|^{-2\Delta_\phi} \; ,
\end{split}
\end{equation}
 and we notice that $C(0,x)$ is (half) the GFFT covariance of section \ref{sec:GFFT}, with the replacement $\z\to\z/2$.

\paragraph{Bilinear operators.}

As the theory is Gaussian and massless, we expect again that in the (non-conformal) OPE of two fundamental fields, $\phi\times \phi$, only bilinear operators in $\phi$ appear, that is, operators of the schematic form:
\begin{equation} \label{BilinearPrimaries}
     [\phi\phi]_{k,l,\m_1 \cdots \m_J}(\t,x)=  \, :\phi(\t,x) (\p_\t)^k(\partial_i\partial_i)^l \left( \partial_{i_1}...\partial_{i_J}-\text{traces} \right)\phi(\t,x) : \; ,
\end{equation}
whose scaling dimension is
\begin{equation}
    \D_{k,l,J}= 2\D_{\phi} + J + 2l + k \z \; .
\end{equation}
Notice that while we have the usual degeneracy for a fixed integer $n$ such that $J+2l=2n$, the time derivatives are in general (for non-rational $\z$) unambiguously identifiable. 
On the other hand, lacking special conformal transformations to select primary operators, the distribution of derivatives on the two fields is rather arbitrary.

\subsection{The massless free fractional Lifshitz field theory at finite temperature}

The massless free FLFT at finite temperature is defined by the action:
 \begin{equation} \label{eq:aniso-freeaction}
    S[\phi]=\int_0^\beta d\tau \int_{\mathbb{R}^d} d^{d}x \left( \frac{1}{2}\phi(\tau, x) (-\p_\tau^2 ) \phi(\tau, x)+ \frac{1}{2}\phi(\tau, x) (-\p_i \p_i )^\z \phi(\tau , x)\right) \; .
\end{equation}
In this case, the covariance, or two-point function, is:
\begin{equation}\label{eq:covtemp}
   C_{\beta}(\tau,x)= \frac{1}{\beta}\sum_{n=-\infty}^{+\infty} \int \frac{d^{d}p}{(2\pi)^{d}} \, \frac{ e^{\imath p \cdot x+ \imath \omega_n \tau}}{p^{2\z}+\omega_n^2} \;,
\end{equation}
where $\omega_n=2 \pi n/\beta$ and the subscript $\beta$ tracks the inverse temperature. In this notation, the covariance at zero temperature of the previous section is denoted $C_{\infty}(\tau,x)$.
After using the Poisson summation formula, $C_{\beta}$ becomes:
\begin{equation}
C_{\beta}(\tau,x)=  \sum_{m=-\infty}^{+\infty} \int \frac{d^{d}p}{(2\pi)^{d}}  \int \frac{d\omega}{(2\pi)} \, \frac{e^{\imath p \cdot x+ \imath \omega\tau_m }}{p^{2\zeta}+\omega^2} 
    \; ,
\end{equation}
where $\tau_m=\tau+\beta m$. Following the same steps as in the previous section we get the massless covariance at finite temperature: 
\begin{equation}
C_{\beta}(\tau,x) = \sum_{m=-\infty}^{+\infty} \int \frac{d^{d}p}{(2\pi)^{d}} e^{\imath p \cdot x } \f{e^{-|\t_m|(p^2)^{\z/2}}}{2 (p^2)^{\z/2}}  \; ,
\end{equation}
which becomes, after summing over $m$ (restricting to the fundamental domain $0\leq \tau <\beta$):\footnote{
For $0\leq \tau<\beta$, we have 
$\sum_m e^{-a|\tau + \beta m|} 
    = e^{-a\tau} +\sum_{m\ge 1} e^{-a\tau - a\beta m}
    + \sum_{m\ge 1} e^{a\tau - a \beta m} =
    e^{-a\tau} + \frac{ e^{a\tau}  + e^{-a\tau }}{e^{a\beta} -1}$ for any $a$ with positive real part. \label{foot:thermal_sum}
} 
\begin{equation}\label{eq:apropape}
C_{\beta}(\tau,x) =\int \frac{d^{d}p}{(2\pi)^{d}} \, e^{\imath p \cdot x } 
\f{1}{2 (p^2)^{\z/2}} \, \left(
 e^{- \tau (p^2)^{\frac{\zeta}{2}} } 
 + \frac{ 
e^{\tau (p^2)^{\frac{\zeta}{2}}   }
+ e^{ - \tau (p^2)^{\frac{\zeta}{2}}  } }
{ e^{ \beta (p^2)^{\frac{\zeta}{2} }   } -1 
} 
\right)   \;.
\end{equation}

The first term in parentheses is the $m=0$ mode and reproduces the zero-temperature result. This is the only surviving term in the $\beta \to \infty$ limit. The second term captures the finite temperature effects. 

As in the case of the GFFT at finite size, the two-point function can be seen as the generating function for the one-point functions of scaling operators. 
For example if we set $x=0$ and Taylor expand in $\tau$ in \eqref{eq:covtemp} we find:
\begin{equation}\label{eq:coefint}
    C_{\beta}(\tau,0) =\sum_{k\ge0} \frac{(-1)^k\tau^{2k}}{(2k)!} \left(\frac{1}{\beta} \sum_{n=-\infty}^{+\infty} \int \frac{d^{d}p}{(2\pi)^{d}} \frac{\omega_n^{2k}}{p^{2\z}+\omega_n^2} \right) \;,
\end{equation}
and the coefficient of $\tau^{2k}$ is the one-point function of the bilinear scaling operator ${\cal O}_{k}(\tau',x) = :\phi \partial_\tau^{2k} \phi (\tau',x):$ at finite temperature.\footnote{Operators with odd number of time derivatives vanish even at finite temperature, due to time-reversal symmetry.}
As the integral in \eqref{eq:coefint} is divergent, one needs to regularize it by an explicit subtraction. This is fairly straightforward in this case: in order to cure all the ultraviolet divergences it is enough to subtract the zero-temperature two-point function,
\begin{equation}
\begin{split}
&  \sum_{k\ge0} \frac{(-1)^k\tau^{2k}}{(2k)!}  \braket{{\cal O}_k(\tau',x)}  =   C_{ \beta}(\tau,0)- C_{\infty}(\tau,0)  \crcr
& \qquad =\sum_{k\ge0} \frac{(-1)^k\tau^{2k}}{(2k)!} \left(\frac{1}{\beta} \sum_{n=-\infty}^{+\infty} \int \frac{d^{d}p}{(2\pi)^{d}} \frac{\omega_n^{2k}}{p^{2\z}+\omega_n^2} - \int \frac{d\omega}{2\pi} \int \frac{d^{d}p}{(2\pi)^{d}} \frac{\omega^{2k}}{p^{2\z}+\omega^2}   \right)   \; ,
\end{split}
\end{equation}
which comes to subtracting the $m=0$ mode after Poisson summation.

This subtraction can be understood directly from the representation~\eqref{eq:apropape}, which we now rewrite in $d=2$ and at $x= 0$, with the change of variables $t=p^2$:
\begin{equation}
C_{\beta}(\tau,0) = \frac{1}{ 8\pi }
    \int_0^{\infty} dt  \;  
    \;t^{-\frac{\zeta}{2}}  \;  \bigg( e^{-\tau t^{\frac{\zeta}{2} } } + \frac{ 
e^{\tau t^{ \frac{\zeta}{2}}  } + e^{ - \tau 
t^{\frac{\zeta}{2}}  } }
{ e^{ \beta  t^{\frac{\zeta}{2} }  }  -1 } \bigg)\;.
\end{equation}
A naive Taylor expansion of the integrand yields ultraviolet (large $t$) divergent terms at every order, coming form the first term (the zero-temperature $m=0$ mode):
\begin{equation}
     \frac{1}{ 8\pi }
    \int_0^{\infty} dt  \;  
     \sum_{k\ge 0} \frac{(-1)^k}{k!} \tau^k \;
      t^{\frac{\zeta}{2} (k-1) } \;.
\end{equation}
However, once they are summed over $k$ they reconstitute $C_{\infty}(\tau,0)$ (compare to \eqref{eq:C-Lifshitz-cases}):
\begin{equation}
      \frac{1}{ 8\pi }
    \int_0^{\infty} dt  \;  
    \;t^{-\frac{\zeta}{2}}  \;  e^{-\tau t^{\frac{\zeta}{2} } } = \tau^{-\frac{2\Delta_\phi}{\z} } \; \frac{\G(2/\z-1)}{ 4\pi \z}
      \equiv \tau^{-\frac{2\Delta_\phi}{\z} } \; \tilde{d}_{ -2\Delta_\phi / \z  } \;,
\end{equation}
 which is unsurprising as the divergence came from expanding in $\tau$ the $m=0$ mode.

Subtracting these terms in the spectral representation and Taylor expanding the second term in $\tau$ we identify the one-point functions of the scaling operators ${\cal O}_k$:
\begin{equation}
   (-1)^k  \braket{{\cal O}_k(\tau',x)}  =  
    \frac{1 }{4\pi} \int_0^{\infty} dt  \;  
    \; \frac{ t^{\zeta (k -\frac{1}{2} ) }  }
{ e^{ \beta  t^{\frac{\zeta}{2} }  } -1 }
= \f{ \G(2\Delta_\phi/\z  +2k) \ \text{Li}_{2\Delta_\phi/\z  +2k}(1)}{2\pi\z\, \beta^{ 2\Delta_\phi/\z  +2k}}  \equiv  \beta^{ -\frac{2\Delta_\phi}{\z}  -2k}\;  d_{2k} 
\; .
\end{equation}
Observe that the counterterm part $C_{\infty}(\tau,0) \sim \tau^{-2\Delta_\phi / \z }  $ is non-analytic at $\tau=0$, that is, the two-point function is:
\begin{equation}
C_{\beta}(\tau,0) = \tau^{ -\frac{2\Delta_\phi}{\z}    } \; \tilde{d}_{ -2\Delta_\phi / \z  } + \sum_{k\ge 0} \frac{ \tau^{2k} }{(2k)!} \; \beta^{ -\frac{2\Delta_\phi}{\z}    -2k} d_{2k} \;, 
\end{equation}
and the subtraction is designed to take out the non-analytic piece of the two-point function, corresponding to the identity operator's contribution to the OPE.

\subsection{The interacting fractional Lifshitz field theory at zero temperature}

We now turn on the quartic interaction in the zero-temperature theory, that is, we consider the full model \eqref{eq:aniso-action} on $\mathbb{R}^{d+1}$, and we are going to show that at large $N$ the theory exhibits an interacting infrared fixed point, much like its isotropic counterpart. 

The Schwinger-Dyson equation at zero temperature
relates the free covariance $\tilde C^{-1}(\omega,p)=\omega^{2}+(p^2)^{\z}$ and the full interacting propagator (two-point function). As the combinatorics is the same as in the usual $O(N)$ model, at large $N$ we have the usual dominance of cactus diagrams, and thus the Schwinger-Dyson equation reads:
\begin{equation} 
\tilde G(\omega_0,p_0)^{-1} = \tilde C(\omega_0,p_0)^{-1} + M_{\text{bare}}^{2\z} +\frac{\lambda}{2}\int \frac{d^{d} p}{(2\pi)^{d}}  \int \frac{d \omega}{2\pi}  \; \tilde G(\omega,p) \, ,
\end{equation}
which as usual gives only a mass correction to the free propagator.
 Like in the isotropic case, we tune the bare mass in order to cancel such mass correction and obtain a massless propagator $\tilde G^{-1}(\omega,p)=\omega^{2}+(p^2)^{\z}$, that is, we impose the condition: 
\begin{equation} \label{eq:aniso-Mbare}
 0=M_{\text{bare}}^{2\zeta} +\frac{\lambda}{2}  \int \frac{d^{d} p}{(2\pi)^{d}} \int \frac{d\omega}{2\pi} \;  \frac{1}{\omega^2 + (p^2)^\z}  \equiv 
M_{\text{bare}}^{2\zeta} +\frac{\lambda}{2}  \cT
\; ,
\end{equation}
where of course a ultraviolet regularization is implicit for the massless tadpole $\cT=C (0,0)$.

In order to show that the model exhibits an infrared fixed point in the large-$N$ limit, we derive the beta function of the quartic coupling. As usual in the vector large-$N$ limit, the beta function is one-loop exact and at one loop only the bubble graph contributes. We regularize the infrared by performing a subtraction at a non-zero external momentum $(\omega_0,0)$ such that $\omega_0=\mu^{\zeta}$.\footnote{Alternatively we can subtract at an external momentum $(0,p)$ such that $p^2=\mu^2$. The computation in this case is slightly more involved but leads to the same result, as expected.} The bubble integral with massless propagator is then:
\begin{equation}
\begin{split}
    B &=\int \frac{d\omega}{2\pi} \int \frac{d^d q}{(2\pi)^d} \frac{1}{(\omega^2+(q^2)^{\zeta})((\omega+\omega_0)^2+(q^2)^{\zeta})} \\
    & =
    \int \frac{d^d q}{(2\pi)^d}\frac{(q^2)^{-\zeta/2}}{4(q^2)^{\zeta}+\omega_0^2}  =-\frac{2^{1-\frac{d}{\zeta}}
     \pi \omega_0^{ \frac{d}{\zeta}-3}
    }{(4\pi)^{\frac{d}{2} } \Gamma(\frac{d}{2} )  \zeta \cos(\frac{\pi d}{2\z})} \, ,
\end{split}
\end{equation}
where we integrated over $\omega$ by deforming the contour and using the residue theorem. In the weakly relevant case $\zeta=\frac{d+\epsilon}{3}$, this integral exhibits a pole $B=\frac{\mu^{-\epsilon} }{2 (4\pi)^{d/2}\Gamma(d/2) \epsilon} $.  The large-$N$ proper vertex at the subtraction point is: 
\begin{equation}
    \Gamma^{(4)}(\omega_0,0)=  \f{2\lambda}{1+ \lambda B }\, ,
\end{equation}
and defining the running coupling $ g=\mu^{-\epsilon}\Gamma^{(4)}(\omega_0,0)/2$, we obtain the large-$N$ beta function:
\begin{equation}
    \beta(g) =\mu \partial_{\mu} g  = -\epsilon g + \frac{g^2}{2(4\pi)^{d/2}\Gamma(d/2)} \,  .
\end{equation}
This beta function displays a non-trivial fixed point 
$ g^{\star}=2 (4\pi)^{d/2}\Gamma(d/2)\epsilon$, corresponding to a bare coupling $\lambda$ flowing to infinity.

\subsection{The interacting fractional Lifshitz field theory at finite temperature}

At last, we turn on also the temperature, that is, we compactify the Euclidean time direction and study the full model \eqref{eq:aniso-action} on $S_\b^1\times\mathbb{R}^{d}$, and we are going to discuss the thermal mass and one-point functions of the scaling operators. 

The model is tuned to the same fixed point, still at large $N$, but the  compactification of time implies that the tadpole is no longer canceled by the bare mass in \eqref{eq:aniso-Mbare}, and therefore we have a solution of the Schwinger-Dyson equation of the form $\tilde G_\b^{-1}(\omega_n,p)=\omega_n^{2}+(p^2)^{\z}+M_{th}^{2\z}$,
with $\omega_n=2 \pi n/\beta$ and non-vanishing thermal mass $M_{th}$. The thermal mass is fixed self-consistently by the mass gap equation:
\begin{equation}
\begin{split}
M_{th}^{2\z} &=M_{\text{bare}}^{2\z} +\frac{\lambda}{2} \; \frac{1}{\beta}\sum_n \int  \frac{d^{d} p}{(2\pi)^{d}} \frac{1}{\omega_n^2 + (p^2)^\z+ M_{th}^{2\z} } \\ 
&= \frac{\lambda}{2}\Bigg( \sum_{m \ne 0}  \int \frac{d^{d} p}{(2 \pi)^{d}} \int \frac{d\omega}{2\pi} \; \frac{e^{\imath \b m \omega}}{\omega^2+(p^2)^{\zeta}+M_{th}^{2\zeta}} \\
&\qquad\qquad  - M_{th}^{2\zeta}\int \frac{d^{d} p}{(2 \pi)^{d}} \int \frac{d\omega}{2\pi} \; \frac{1 }{(\omega^2 +(p^{2})^{\zeta}+M_{th}^{2\zeta})(\omega^2 +(p^{2})^{\zeta})}  \Bigg) \\
& \equiv \frac{\lambda}{2}\, \left( \cT(M_{th}) -\cT \right) \; ,
\end{split}
\end{equation}
where in the second equality we used the Poisson summation formula \eqref{eq:Poisson}, and we have combined the $m=0$ term with the explicit expression of the bare mass. In the last step we have defined the subtracted thermal tadpole $ \cT(M_{th}) -\cT =G_\b(0,0) - C_{\infty}(0,0)$.

At the fixed point, the bare coupling $\lambda$ diverges, and thus the mass gap equation becomes:
\begin{equation}
    \cT(M_{th}) -\cT = 0 \;.
\end{equation}
We are going to show that it admits a unique non-vanishing solution for $\z\in (d/3,1)$.

First, we write the two-point function at finite temperature in position space in the Fourier representation:
\begin{equation}\label{eq:MaxGDef}
 G_{\beta}(\tau,x)= \frac{1}{\beta}\sum_{n=-\infty}^{+\infty} \int \frac{d^{d}p}{(2\pi)^{d}}  \; \frac{e^{\imath p \cdot x+ \imath \omega_n \tau}  }{p^{2\z}+\omega_n^2  + M_{th}^{2\z}  } = \sum_{m=-\infty}^{+\infty} \int \frac{d^{d}p}{(2\pi)^{d}} \int \frac{d\omega}{2\pi} \frac{e^{\imath p \cdot x + \imath \omega \tau_m } }{p^{2\zeta}+\omega^2 + M_{th}^{2\z}} \; , 
\end{equation}
with $\tau_m=\tau+\beta m$. 
The integral over $\omega$ is evaluated again by the residue theorem and restricting to the fundamental domain $\t\in[0,\b)$ the sum over $m$ is just a geometric series (see footnote~\ref{foot:thermal_sum}), leading to:
\begin{equation}
   G_{\beta}(\t,x) = \int \frac{d^{d}p}
   {(2\pi)^{d}} e^{\imath p \cdot x } 
  \f{1} {2 \sqrt{ M_{th}^{2\z} + (p^2)^{\z} }} \bigg( e^{-\t \sqrt{ M_{th}^{2\z} + (p^2)^{\z} } }  + \frac{e^{\tau \sqrt{ M_{th}^{2\z} + (p^2)^{\z} }}  + 
  e^{-\tau \sqrt{ M_{th}^{2\z} + (p^2)^{\z} }}
  }{e^{\beta \sqrt{ M_{th}^{2\z} + (p^2)^{\z} } }-1}\bigg)\; ,
\end{equation}
where the first term in parentheses is the $m=0$ term.

\begin{figure}[htbp]
\centering
\includegraphics[scale=0.70]
{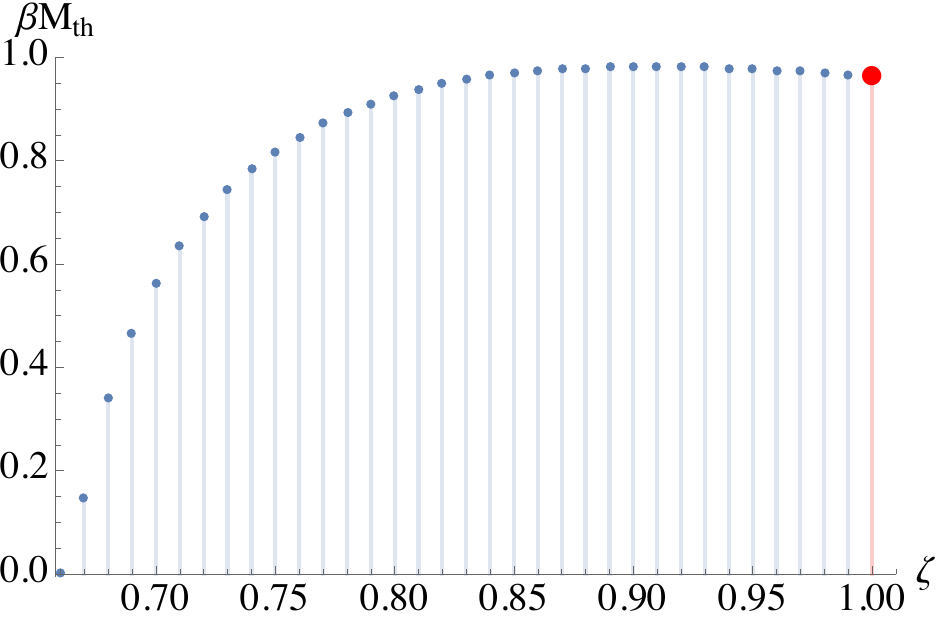}
\caption{Numerical solutions for the thermal mass as a function of $\z$ in $2+1$ dimensions. The red dot is the value of the short-range thermal mass, and it is approached in the $\z\rightarrow1$ limit, while the solution goes to zero for $\z\to 2/3$. 
}
\label{fig:TheMass1}
\end{figure}

Setting $d=2$, taking $\t=x=0$, and changing variables to $p^2=t$, we arrive at the following integral form of the mass gap equation:
\begin{equation}
\cT(M_{th}) -\cT =     \frac{1}{8\pi}   \int_0^{\infty} dt \; 
    \left[ \frac{ 1 }
{ \sqrt{  M_{th}^{2\zeta} + t^{\zeta } } } 
 \left( 1  
  + \frac{ 2 }{  e^{\beta \sqrt{  M_{th}^{2\zeta} + t^{\zeta }   }  }  -1}
 \right) - \frac{1}{\sqrt{t^{\zeta}}}   \right] =0 \; ,
\end{equation}
that always admits a unique solution.\footnote{We rewrite the mass gap equation as $f_1(M_{th}) = f_2(M_{th})$ with:
\begin{equation}
f_1(M_{th}) =   \int_0^{\infty} dt \; 
    \frac{ 1 }
{ \sqrt{  M_{th}^{2\z} + t^{\zeta } } } 
 \left( 
 \frac{ 2 }{  e^{\beta \sqrt{  M_{th}^{2\z} + t^{\zeta }   }  }  -1}
 \right) \;,\qquad 
 f_2(M_{th}) =   \int_0^{\infty} dt \; 
    \left[ \frac{1}{\sqrt{t^{\zeta}}}   - \frac{ 1 }
{  \sqrt{  M_{th}^{2\z}+ t^{\zeta } } }  \right]  \; ,
\end{equation}
and $f_2$ is convergent for $\zeta >\frac{2}{3}$ as $f_2\sim \int^{\infty} dt \; t^{-3\z/2}$. 
We note that $f_1(0) >0$ and $f_1$ is strictly decreasing with $M_{th}$ while $f_2(0) = 0$ and $f_2$ is strictly increasing with  $M_{th}$.}
 We have plotted a numerical solution as a function of $\z$ in Figure~\ref{fig:TheMass1}.

Keeping $\t\neq 0$, we obtain instead:
\begin{equation}\label{eq:MaxGeval}
    G_{\beta} (\tau, 0)= \frac{1}{8\pi}
    \int_0^{\infty} dt \;
 \frac{ 1 }
{ \sqrt{  M_{th}^{2\z}+ t^{\zeta } } } 
 \left( e^{-\tau \sqrt{  M_{th}^{2\z}+ t^{\zeta }  }  }  
  + \frac{ e^{ \tau \sqrt{  M_{th}^{2\z}+ t^{\zeta } }  } + e^{-\tau \sqrt{  M_{th}^{2\z}+ t^{\zeta }  }  }  }{  e^{\beta \sqrt{  M_{th}^{2\z}+ t^{\zeta }   }  }  -1}
 \right) \; ,
\end{equation}
and we are interested in its small $\tau$ expansion, which again we relate to the one-point function of the bilinear scaling operator ${\cal O}_{k}(\tau',x) = :\phi \partial_\tau^{2k} \phi (\tau',x):$ at finite temperature:
\begin{equation} \label{eq:aniso-generFunct}
     \sum_{k\ge0} \frac{(-1)^k\tau^{2k}}{(2k)!}  \braket{{\cal O}_k(\tau',x)}  = G_{\beta} (\tau, 0) - \text{counterterms} \;.
\end{equation}
As in the isotropic case, the $k=0$ operator will actually drop from this equation, as it is not in the spectrum of the interacting fixed point, where it is replaced by the intermediate field $\s$.
Moreover, as in section \ref{sec:higher-twist_int}, the counterterms will contain useful information about other operators in the spectrum.

The need for counterterms comes from the first term in parentheses in \eqref{eq:MaxGeval}, i.e.\ the $m=0$ mode:
its naive Taylor expansion in $\tau$ exhibits ultraviolet divergent terms at each order. Such divergences are renormalized by subtracting the beginning of the Taylor expansion in $M^{2\z}$ up to the appropriate order:
\begin{equation}
  \frac{1}{8\pi}
    \int_0^{\infty} dt \;
    \sum_{q\ge 0} \frac{(-\tau)^q}{q!} \bigg[ (M_{th}^{2\zeta} + t^{\zeta} )^{\frac{q-1}{2}} 
    -\sum_{n=0}^{ \lfloor \frac{q}{2} \rfloor } 
     \frac{ \Gamma(\frac{q+1}{2} ) }{n! \Gamma( \frac{q+1}{2} -n ) }(M_{th}^{2\zeta} )^n t^{ \zeta (\frac{q-1}{2}  -n )}\bigg]
\;.
\end{equation}
The odd $q=2k+1$ terms are entirely subtracted, since $(M_{th}^{2\zeta} + t^{\zeta} )^k$ is a polynomial in $M^{2\z}$, while for even $q=2k$ only the terms containing a power of $t$ larger than $-1$ are subtracted:
\begin{equation}
  \frac{1}{8\pi}\frac{\tau^{2k}}{(2k)!}
    \int_0^{\infty} dt \;
      \bigg[ (M_{th}^{2\zeta} + t^{\zeta} )^{k- \frac{1}{2}} -\sum_{n=0}^{ k } 
     \frac{ \Gamma( k + \frac{1}{2} ) }{n! \Gamma( k-n+ \frac{1}{2} ) }(M_{th}^{2\zeta} )^n t^{ \zeta (k -n -\frac{1}{2})}\bigg]  \;, 
\end{equation}
resulting in an expression that is both ultraviolet ($t\to \infty$) and infrared ($t\to 0$) convergent for $\z > 2/3$.
 Indeed, convergence in the infrared is obvious term by term; convergence in the ultraviolet follows by noting that the subtracted integrand behaves like $t^{-\frac{3}{2} \zeta}$ at large $t$. 
Grouping the counterterms by their power in $M_{th}$, we obtain the following small-$\tau$ expansion of the massive two-point function in finite temperature (recall that $2\Delta_\phi=2-\zeta$):
\begin{align}\label{eq:OPEgood}
    G_{\beta}(\tau,0)  = \tau^{- \frac{2\Delta_\phi}{\z} } \left (\sum_{n=0}^\infty (\t M_{th}^\z)^{2 n } \tilde{d}_{ - 2\Delta_\phi/\z + 2n } + \sum_{k\ge 0} \frac{ (\t M_{th}^\z)^{2k+\frac{2\Delta_\phi}{\z}} }{(2k)!} d_{2k}^{(\beta M_{th}^\z)} \right) \; ,
\end{align}
exhibiting the finite-temperature scaling behavior 
$ G_{\beta}(\tau,0)  = \tau^{- 2\D_\phi / \z  }\, h\left(  \t M_{th}^\z \right)$
for a function $h(u)$ that is finite, but non-analytic, at $u=0$.
Taking into account the prefactor, the non-analytic part in $\t$ of $G_{\beta}(\tau,0)$ is the first term in parentheses in \eqref{eq:OPEgood}, while the second term gives the analytic part.

The coefficients of the non-analytic part of $G_{\beta}(\tau,0) $ are: 
\begin{equation}\label{eq:nonana}
\begin{split}
  \tilde{d}_{ - 2\Delta_\phi/\z + 2n }  &=   \int_0^{\infty} \frac{dt}{8\pi} \; \sum_{q\ge 2n} 
   \frac{(-1)^q}{q!}    \;
     \frac{ \Gamma(\frac{q+1}{2} ) } {n! \Gamma( \frac{q+1}{2} -n ) } \, t^{
    \zeta (\frac{q-1}{2}  -n )} \\
    &=  \int_0^{\infty} \frac{dt}{8\pi} \;\left( \frac{ _1F_2\left(1;\frac{1}{2},n+1;\frac{t^\z}{4}\right)}{4^{n} \,t^{\z/2}
   (n!)^2}-\frac{\, _0F_1\left(;n+\frac{3}{2};\frac{t^\z}{4}\right)}{ (2 n+1)!} \right) \;,
\end{split}
\end{equation}
and, although each term at $q$ fixed is ultraviolet divergent, performing the sum over $q$ first we obtain a convergent integral expression due to the alternating sign (the linear combination of hypergeometric functions decays like $t^{-\z(n+1)/2} e^{-t^{\z/2}}$ for $t\to+\infty$).

The analytic part of $G(\t,0)$ comprises effects due to both the finite temperature and the presence of a non-zero thermal mass:
\begin{equation}
\begin{split}
  d_{2k}^{(\beta M_{th}^\z)}  = &  \int_0^{\infty} \frac{dt}{8\pi} \;
  \bigg[ (1 + t^{\zeta} )^{k- \frac{1}{2}} -\sum_{n=0}^{ k } 
     \frac{ \Gamma( k + \frac{1}{2} ) }{n! \Gamma( k-n+ \frac{1}{2} ) } t^{
    \zeta (k  -n -\frac{1}{2})} +  \frac{ 2 \left(  1+ t^{\zeta }  \right)^{k-\frac{1}{2} }  } 
{e^{\beta M_{th}^\z \sqrt{  1+ t^{\zeta }   }  }  -1  }  \bigg]  \\
= &  \int_0^{\infty} \frac{dt}{8\pi} \;
  \bigg[ -\frac{\Gamma \left(k+\frac{1}{2}\right) \,
   _2{F}_1\left(1,\frac{3}{2};k+2;-t^{-\z}\right)}{2 \sqrt{\pi } \, (k+1)!\, t^{3\z/2}}  +  \frac{ 2 \left(  1+ t^{\zeta }  \right)^{k-\frac{1}{2} }  } 
{e^{\beta M_{th}^\z \sqrt{  1+ t^{\zeta }   }  }  -1  }  \bigg] \;,
\end{split}
\end{equation}
and we stress again that this is a convergent integral for $\zeta>2/3$.

As in the CFT case, the coefficients of the small-$\t$ expansion can be related to the one-point functions of composite operators.
Indeed, from \eqref{eq:aniso-OPE} and \eqref{eq:aniso-OPEcoeff}, we have:
\begin{equation}\label{eq:genOPE}
 \phi(\tau, 0) \phi(0,0)  = 
  \sum_{\cal O} f_{\phi\phi {\cal O} }(0) \; |\tau|^{\frac{\Delta_{\cal O}  - 2\Delta_\phi }{\zeta }} \;  {\cal O}(0,0) \;,
\end{equation}
where ${\cal O}$ runs over all the scaling operators in the theory. 
Taking the expectation value of \eqref{eq:genOPE}, we conclude that the two-point function is a generating function for the one-point functions of the operators ${\cal O}$ having $f_{\phi\phi {\cal O} }(0)\neq 0$.
Notice that the operators in \eqref{eq:genOPE} are not primaries, as we do not have conformal invariance here, hence they include also total derivatives of composite operators. However, the latter have vanishing one-point functions even at finite temperature, due to translation invariance.

Seen from such an OPE perspective, we recognize that
\eqref{eq:OPEgood} is indexed by two classes of operators. First, we have 
operators with momentum dimensions $\frac{ \Delta_{\cal O} -2\Delta_{\phi}  }{\zeta}= 2k$, i.e.\ $\Delta_{\cal O} = 2\Delta_\phi + 2k\zeta $. These are identified with the operators ${\cal O}_k(\tau,x) = :\phi\partial_\tau^{2k} \phi(\tau,x):$. By consistency with \eqref{eq:aniso-generFunct}, we must have that with our choices of normalizations $f_{\phi\phi {\cal O} }(0)=(-1)^k/(2k)!$,
and thus we conclude that:
\begin{equation}
   (-1)^k  \braket{{\cal O}_k(\tau',x)}  = M_{th}^{2\D_\phi+2k\z}  d_{2k}^{(\beta M_{th}^\z)} \; .
\end{equation}
Notice that the mass gap equation corresponds to $d_0^{(\beta M_{th})} = 0$.

Second, we have a family of operators with momentum dimension $\frac{ \Delta_{\cal O} -2\Delta_{\phi}  }{\zeta}=2n  -\frac{2\Delta_{\phi}}{\zeta} $, i.e.\ $ \Delta_{\cal O}  = 2n\zeta$. As in the isotropic case, the intermediate field $\sigma(x)$, which can be introduced as in \eqref{eq:ON-action-mixed}, 
is identified with the shadow of $\phi^2(x)$ and has momentum dimension $\Delta_{\sigma} = d + \zeta -2\Delta_{\phi} = 2\zeta$.  Therefore, the second class of operators contributing to the OPE of the two-point function correspond to composite $\sigma^{n}$ operators:
\begin{equation}
    f_{\phi\phi \s^n }(0) \braket{\s^n(\t',x)} =  M_{th}^{2n\z} \  \tilde{d}_{-2\D_\phi/\z + 2n } \; .
\end{equation}

\section{Conclusions}

We have studied classical and quantum versions of the long-range $O(N)$ model at large $N$, focusing on the effects of having one compact spatial direction in the former or finite temperature in the latter.
The quantum model is mapped as usual to a Euclidean field theory with one extra (time) dimension, along which the action is local, thus resulting in an anisotropic action, which at criticality has a Lifshitz scale invariance, but no conformal invariance. This should be contrasted to the short-range case, where the classical and quantum model are described by the same field theory, in $d$ and $d+1$ dimensions, respectively, and therefore they are both conformal at criticality.
 Therefore, in the long-range model also the finite-size and finite-temperature compactifications are quite different: the finite-size geometry corresponds to compactifying a direction along which the model has long-range interaction, while the finite temperature corresponds to compactifying the local Euclidean time direction.

In both cases, the non-compact versions of the model exhibit an ultraviolet Gaussian fixed point and an interacting infrared fixed point. Upon compactification, the fixed-point data (conformal data in the classical case, or simply fixed-point data in the quantum one) of the non-compact theories needs to be supplemented with the non-zero values of one-point functions of primary (respectively scaling) operators. 

We have focused on the operators arising in the OPE of two fundamental fields. The corresponding new data can be inferred from the small argument behaviour of the two-point 
function, and consequently our main technical tool consists in finding suitable expansions of the latter. Our results are:
\begin{itemize}
    \item for the free (ultraviolet) compactified theories, we obtained the one-point functions of bilinear operators. In the finite-size case we obtained such data for all the bilinear primaries, with arbitrary spin and twist. In the finite-temperature case we computed this data only for scalar bilinears with time derivatives.
    
    \item in the case of the compactified interacting (infrared) theories, at large $N$, the effect of the interaction is encoded in the appearance of a dynamically generated mass scale which we determined self-consistently. 
    We obtained the fixed-point data for the same set of bilinear operators as in the free case.
    Moreover, in the interacting cases, a second infinite family of operators contributes to the OPE, consisting in arbitrary powers of the Hubbard--Stratonovich intermediate field. We computed also the one-point functions of such operators.
\end{itemize}

A natural extension of this work would be to consider other configurations of the background geometry, for example considering the classical model with more or all directions compactified, or the quantum model with compact spatial directions, either at zero or finite temperature.

Another important future direction would be the inclusion of subleading corrections in $1/N$.
In the short-range case, the effect of next-to-leading corrections on the new finite-temperature data has been studied in \cite{Diatlyk:2023msc}. In the long-range case, we expect further complications. In the classical isotropic case, we have seen that even in the leading-order we considered here, the non-standard propagator makes analytic computations much harder, and certainly this will affect the loop integrals that are needed for subleading orders. In the anisotropic (quantum) case, computations are harder even at zero temperature; in fact, as we mentioned in the introduction, the fractional Lifshitz field theory we considered here has not been studied much before, and there are several open directions even at zero temperature. 
In particular, it would be interesting to determine the deviation of the dynamic critical exponent $z$, characterizing the Lifshitz scaling at the fixed point, from its leading order value $z=\z$, and to ascertain the existence of a crossover (or transition) to isotropic short-range behavior above some critical value of $\z$. Such aspects of the zero-temperature FLFT require at least the next-to-leading order in $1/N$, therefore we postpone them to future work.

\section*{Acknowledgments}
Nordita is supported in part by NordForsk. R. G. and D. L. have been supported by the European Research Council (ERC) under the European Union's Horizon 2020 research and innovation program (grant agreement No 818066) and by the Deutsche
Forschungsgemeinschaft (DFG) under Germany's Excellence Strategy EXC--
2181/1 -- 390900948 (the Heidelberg STRUCTURES Cluster of Excellence).

\newpage
\appendix

\section{Minimal twist bilinear operators }
\label{app:bilinear}

The most general bilinear primary operator with spin $J$ is generated by acting with a precise combination of derivatives on the two fields. The combination is fixed by requiring that the operator is annihilated by the generator of special conformal transformations.
The result of imposing such constraint on minimal twist operators was obtained in \cite{Dobrev:1975ru,Craigie:1983fb} (see also \cite{Skvortsov:2015pea,Giombi:2016ejx}), and it is most easily expressed by introducing a complex polarization vector $\xi^\m$ such that $\xi^2=0$ with which to form a scalar:
\begin{equation} \label{eq:spin-Op}
\begin{split}
\cO_{0,J}(x,\xi) &\equiv [\phi_a\phi_a ]_{0,\m_1 \cdots \m_J} (x) \xi^{\m_1} \cdots \xi^{\m_J} = \f{1}{N} :  \phi_a(x)f_J\left(\xi\cdot\pleft,\xi\cdot\pright\right) \phi_a(x) : \; , 
\end{split} 
\end{equation} 
where the colon notation stands as usual for  normal (or Wick) ordering (e.g.\ \cite{salmhofer:book}), which serves to cancel the divergence in the one-point function. The function $f_J(u,v)$ is found to be
\begin{equation} \label{eq:f-Gegenbauer}
\begin{split}
    f_J(u,v) &= \f{J!\, \G\left(\n^{(\z)}\right)}{4^J\, \G\left(\n^{(\z)}+J\right)} \left(u+v \right)^J C^{\nu^{(\zeta)}}_J\left(\f{u-v}{u+v}\right)= \sum_{n=0}^{J/2} \g_{n,J} \left(u+v\right)^{2n} \left(u-v \right)^{J-2n} \; ,
\end{split}
\end{equation}
with $C^{\nu^{(\zeta)}}_J(z)$ being a Gegenbauer polynomial with $\nu^{(\zeta)}=\frac{d-2\zeta-1}{2}$, and
\begin{equation}
    \g_{n,J} = (-1)^n \f{J!\, }{2^J\, \G\left(\nu^{(\zeta)}+J\right)} \f{\G\left(J+\nu^{(\zeta)}-n\right)}{2^{2n}\, n! (J-2n)!}  \; .
\end{equation} 

The normalization is chosen for convenience as in  \cite{Craigie:1983fb} (in particular it gives finite coefficients for $J>0$ in $d=3$).

\subsection{Normalization of two-point functions}
The two-point function at zero temperature can be computed in the free theory by applying the above formula to the (formal) spin-0 case \cite{Craigie:1983fb}
\begin{equation}
   \la \cO_{0,J}(x_1,\xi_1) \cO_{0,J}(x_2,\xi_2)  \ra
   = \f{2}{N} C(x_{12}) f_J\left(\xi_1\cdot\pleft_1,\xi_1\cdot\pright_1\right) f_J\left(\xi_2\cdot\pleft_2,\xi_2\cdot\pright_2\right) C(x_{12}) \; .
\end{equation}
In order to extract the normalization constant $c_{0,J}$ in 
\begin{equation}
   \la \cO_{0,J}(x,\xi_1) \cO_{0,J}(0,\xi_2)  \ra
   =  c_{0,J} \f{ \left(\xi_1^\m \left( \d_{\m\n} - 2\f{x_\m x_\n}{x^2} \right) \xi_2^\n \right)^J }{(x^2)^{d-2+J}} \; ,
\end{equation}
it is enough to choose $\xi_1=\xi_2=\xi$, in which case the Kronecker delta terms drop out.
The derivatives can be evaluated explicitly using the representation 
\begin{equation}
    C(x) = \frac{\Gamma(d/2-\zeta)}{2^{2\zeta} \pi^{d/2} \Gamma(\z) } \frac{1}{\left(x^2\right)^{d/2-\zeta}}
    = \frac{1}{2^{2\zeta} \pi^{d/2}\Gamma(\z) } \int_0^{+\infty} d\a\, \a^{d/2-\zeta-1} e^{-\a x^2} \; ,
\end{equation}
and the differential formula
\begin{equation}
    (\xi\cdot \p )^k e^{-\a x^2} = (-2\,\a\, \xi\cdot x)^k e^{-\a x^2} \;  .
\end{equation}
Noticing that $\p_1 C(x_{12})=-\p_2 C(x_{12})$ and remembering that $J$ is even, we find 
\begin{equation}
\begin{split}
    &\la \cO_{0,J}(x,\xi) \cO_{0,J}(0,\xi)  \ra
   = \f{2}{N} \left( \frac{1}{2^{2\z} \pi^{d/2} \Gamma(\z) } \right)^2 \int_0^{+\infty} d\a_1\, \a_1^{d/2-\zeta-1} \int_0^{+\infty} d\a_2\, \a_2^{d/2-\zeta-1} \\
   &\qquad\quad\qquad\qquad \times f_J(-2\,\a_1\, \xi\cdot x,-2\,\a_2\, \xi\cdot x)^2 e^{-(\a_1+\a_2) x^2}\\
   &\qquad\quad=\f{2}{N} \left( \frac{1}{2^{2\z} \pi^{d/2}\Gamma(\z) } \f{J!\, \G\left(\n^{(\zeta)}\right)}{4^J\, \G\left(\n^{(\z)}+J\right)} \right)^2 \f{(-2\, \xi\cdot x)^{2J}}{ (x^2)^{d-2\zeta+2J}} \int_0^{+\infty} d\a_1\, \a_1^{d/2-\zeta-1} \int_0^{+\infty} d\a_2\, \a_2^{d/2-\zeta-1} \\
   &\qquad\quad\qquad\qquad \times \left(\a_1+\a_2 \right)^{2J} C^{\n^{(\z)}}_J\left(\f{\a_1-\a_2}{\a_1+\a_2}\right)^2 e^{-(\a_1+\a_2) }
   \\
   &\qquad\quad=\f{2}{N} \left( \frac{1}{2^{2\z} \pi^{d/2}\Gamma(\z) } \f{J!\, \G\left(\n^{(\zeta)}\right)}{4^J\, \G\left(\n^{(\z)}+J\right)} \right)^2 \f{(-2\, \xi\cdot x)^{2J}}{ (x^2)^{d-2\zeta+2J}} 2^{2\zeta+1-d} \int_0^{+\infty} ds\,  s^{d-2\zeta+2J-1} \,e^{-s }\\
   &\qquad\quad\qquad\qquad \times  \int_{-1}^{+1} dt\,(1-t^2)^{\n^{(\z)}-\frac{1}{2}} C^{\n^{(\z)}}_J\left(t\right)^2 \\
   &\qquad\quad=\f{2}{N} \frac{\G(d-2\zeta+2J)}{2^{4\z} \pi^{d} \Gamma(\z)^2 } \left(  \f{J!\, \G\left(\n^{(\z)}\right)}{4^J\, \G\left(\n^{\z}+J\right)} \right)^2 \, 2^{J}\f{\pi 2^{1-4\n^{(\z)}} \G(J+2\n^{(\z)})}{J! (J+\n^{(\z)}) \G(\n^{(\z)})^2}\, \f{(2\, (\xi\cdot x)^2)^{J}}{ (x^2)^{d-2\zeta+2J}} \; ,
\end{split}
\end{equation}
where we used the change of variables $\a_1=s(1+t)/2$, $\a_2=s(1-t)/2$, and the
orthogonality of Gegenbauer polynomials 
\begin{equation}
    \int_{-1}^{+1} dt\,(1-t^2)^{\alpha-1/2} C^{\alpha}_J\left(t\right)^2 = \f{\pi 2^{1-2\alpha} \G(J+2\alpha)}{J! (J+\alpha) \G(\alpha)^2} \; .
\end{equation}

We thus find
\begin{equation}
\begin{split}
    c_{0,J}&=\f{2}{N} \frac{\G(d-2\zeta+2J)}{2^{4\z} \pi^{d} \Gamma(\z)^2 } \left(  \f{J!\, \G\left(\n^{(\z)}\right)}{4^J\, \G\left(\n^{(\z)}+J\right)} \right)^2 \, \f{\pi 2^{1-4\n^{(\z)}+J} \G(J+2\n^{(\z)})}{J! (J+\n^{(\z)}) \G(\n^{(\z)})^2}\\
    &=\frac{2}{N} \frac{\pi ^{\frac{1}{2}-d} J! 2^{-d-2 \zeta -J+2} \Gamma (d+J-2 \zeta -1) \Gamma \left(\frac{d}{2}+J-\zeta \right)}{\Gamma ( \zeta )^2 \Gamma \left(\frac{d}{2}+J-\zeta -\frac{1}{2}\right)} \; .
\end{split}
\end{equation}

This result agrees in the limit $\zeta=1$ with \cite{Skvortsov:2015pea} up to the $\left(  \f{J!\, \G\left(\n^{(1)}\right)}{4^J\, \G\left(\n^{(1)}+J\right)} \right)^2$ factor from the normalization of  \cite{Craigie:1983fb}.

\subsection{OPE coefficient $f_{\phi \phi J}$ }
Now we compute the OPE coefficient between a spinning operator of the kind $\mathcal{O}_{0,J}$ and two fundamental fields. The functional form of the three-point function is constrained by conformal symmetry to be
\begin{equation}\label{ConfStruc3pt}
    \langle \mathcal{O}_{0,J}(x_0,\xi) \phi_i(x_1) \phi_j(x_2) \rangle=\delta_{ij} f_{\phi \phi J}\left(\frac{\xi \cdot x_{20}}{x_{20}^2}-\frac{\xi \cdot x_{10}}{x_{10}^2}\right)^J \frac{1}{(x_{02}^2)^{\frac{d}{2}-\z}(x_{01}^2)^{\frac{d}{2}-\z}} \; .
\end{equation}
We can compute the three-point function explicitly by Wick theorem as we did for the two-point function and extract the OPE coefficient by comparing with \eqref{ConfStruc3pt}. We have
\begin{equation}
\begin{split}
    \langle \mathcal{O}_{0,J}(x_0,\xi) \phi_i(x_1) \phi_i(x_2)\rangle=& \f{2}{N} \left( \frac{1}{2^{2\z} \pi^{d/2} \Gamma(\z) } \right)^2 \int_0^{+\infty} d\a_1\, \a_1^{d/2-\zeta-1} \int_0^{+\infty} d\a_2\, \a_2^{d/2-\zeta-1}\\
   &\qquad \times f_J(-2\,\a_1\, \xi\cdot (x_0-x_1),-2\,\a_2\, \xi\cdot (x_0-x_2)) e^{-\a_1 x_{01}^2-\a_2 x_{02}^2} \; .
\end{split}
\end{equation}
Without loss of generality we can set $x_0=0$ and we get to
\begin{equation}
    \begin{split}
       & \langle \mathcal{O}_{0,J}(0,\xi) \phi_i(x_1) \phi_i(x_2)\rangle= \f{2}{N} \frac{J!\, \Gamma(\n^{(\z)})}{4^{2\z+J} \pi^{d} \Gamma(\z)^2 \Gamma(\n^{(\z)}+J)} \int_0^{+\infty} d\a_1\, \a_1^{d/2-\zeta-1} \int_0^{+\infty} d\a_2\, \a_2^{d/2-\zeta-1}\\
        & \qquad \qquad \times \left(2 \a_1 \frac{\xi \cdot x_1}{x_1^2} +2 \a_2 \frac{\xi \cdot x_2}{x_2^2}\right)^J C^{(\n^{(\z)})}_J\left(\frac{2 \a_1 \frac{\xi \cdot x_1}{x_1^2} +2 \a_2 \frac{\xi \cdot x_2}{x_2^2}}{2 \a_1 \frac{\xi \cdot x_1}{x_1^2} -2 \a_2 \frac{\xi \cdot x_2}{x_2^2}}\right) \frac{e^{-\a_1-\a_2}}{(x_1^2)^{d/2-\z}(x_2^2)^{d/2-\z}} \; .
    \end{split}
\end{equation}
Considering the region $x_2^2\gg x_1^2$ the argument the argument of the Gegenbauer polynomials become $1$ and it is easy to extract
\begin{equation}
    \begin{split}
        f_{\phi \phi J}&=\frac{2}{N} \frac{J!\, \Gamma(\n^{(\z)})}{4^{2\z+J} \pi^{d} \Gamma(\z)^2 \Gamma(\n^{(\z)}+J)} 2^J \Gamma\left(\frac{d}{2}-\z\right) \Gamma\left(\frac{d}{2}-\z+J\right) \frac{\Gamma(2 \n^{(\z)}+J)}{\Gamma(2 \n^{(\z)})J!}\\
        &=\frac{2}{N} \frac{ \Gamma \left(\frac{d -1}{2} -\z \right) \Gamma \left(\frac{d}{2}-\zeta \right) \Gamma (d+J-2 \zeta -1) \Gamma \left(\frac{d}{2}+J-\zeta \right)}{\pi^{d} 2^{4 \zeta +J} \Gamma ( \zeta )^2 \Gamma (d-2 \zeta -1) \Gamma \left(J+\frac{d -1}{2} -\z\right)} \; .
    \end{split}
\end{equation}
Now it is simple to check by a direct computation that in our conventions
\begin{equation}
    \frac{f_{\phi \phi J}}{c_{0,J}}= \frac{1}{J!} \; .
\end{equation}

\bigskip
\bigskip

\newpage

\bibliographystyle{JHEP}
\bibliography{Refs} 
\addcontentsline{toc}{section}{References}

\cleardoublepage

 \end{document}